\documentclass[12pt]{article}
\usepackage{epsf,amsfonts,amssymb,epsfig,amsmath,mathtools,graphics,slashed}
\usepackage{hyperref,hep,graphicx,subfig}
\usepackage{rotating}
\usepackage{xcolor}
\usepackage{verbatim}
\usepackage[final]{showlabels}

\definecolor{greenish}{RGB}{0,190,0}

\definecolor{yellowish}{RGB}{190,190,0}

\definecolor{bluish}{RGB}{0,0,190}

\newcommand{\nn}{\notag \\}

\makeatletter

\@addtoreset{equation}{section}
\makeatother

\begin{document}

\begin{titlepage}

\vfill


\vfill

\begin{center}
   \baselineskip=16pt
   {\Large\bf Dissipative effects in finite density holographic superfluids}
  \vskip 1.5cm
  \vskip 1.5cm
      Aristomenis Donos and Polydoros Kailidis\\
   \vskip .6cm
   \begin{small}
      \textit{Centre for Particle Theory and Department of Mathematical Sciences,\\ Durham University, Durham, DH1 3LE, U.K.}
   \end{small}\\            
\end{center}

\vfill

\begin{center}
\textbf{Abstract}
\end{center}
\begin{quote}
We derive the leading dissipative corrections of holographic superfluids at finite temperature and chemical potential by employing our recently developed techniques to study dissipative effects in the hydrodynamic limit of holographic theories. As part of our results, we express the incoherent conductivity, the shear and the three bulk viscosities in terms of thermodynamics and the black hole horizon data of the dual bulk geometries. We use our results to show that all three bulk viscosities exhibit singular behaviour close to the critical point.
\end{quote}

\vfill

\end{titlepage}

\setcounter{equation}{0}

\section{Introduction}
Holography provides large classes of examples of strongly coupled field theories where exact computations can be carried out \cite{Aharony:1999ti,Witten:1998qj}. In a certain limit, non-trivial questions about field theory can be mapped to well defined problems in Einstein's classical theory of gravity in an asymptotically Anti de-Sitter spacetime (AdS) of dimensionality larger by one. The fall-off conditions of the classical gravitational fields close to the conformal boundary of AdS set the sources of local operators on the field theory side. In the classical limit, the holographic principle states that the partition functions of the two sides are equal, providing a powerful tool to compute expectation values of local operators.

In thermal equilibrium, the geometric dual of the thermal state is a black hole geometry with the temperature set by the Hawking temperature of the Killing horizon. Moreover, the holographic dictionary suggests that global $U(1)$ symmetries on the field theory side are gauged in the bulk. The asymptotic flux of the corresponding gauge fields set the electric charge density of the field theory making holography an invaluable laboratory to study large classes of strongly coupled systems at finite temperature and number density \cite{Hartnoll:2016apf}.

An interesting application of holography concerns the study of systems exhibiting spontaneous breaking of a global symmetry. In general, continuous phase transitions are driven via perturbative instabilities of black holes against fluctuations of classical bulk fields. The first examples of symmetry breaking were due to bulk fields which are charged under continuous internal symmetry groups, leading to superfluid phases of holographic matter \cite{Hartnoll:2008vx,Gubser:2008px}. Spontaneous breaking of spacetime symmetries were realised later in \cite{Nakamura:2009tf,Donos:2011bh,Donos:2011ff,Donos:2011qt} making holography even more appealing for applications in condensed matter systems.

Holographic systems reach local thermal equilibrium with their long wavelength excitations obeying the laws of hydrodynamics. This has been a very active area of research over the past years \cite{Son:2007vk,Bhattacharyya:2008jc,Haack:2008cp,Erdmenger:2008rm,Banerjee:2008th} from which new lessons about low energy effective field theory have been learned \cite{Hartnoll:2016apf,Baggioli:2022pyb}. The hydrodynamic limit of broken symmetry phases where the standard hydrodynamic degrees of freedom of charged fluids combine with gapless Goldstone modes has also been considered in holography \cite{Herzog:2011ec,Bhattacharya:2011tra,Bhattacharya:2011eea}. More recently, the amplitude mode which becomes gapless at continuous phase transitions has been realised holographically in \cite{Donos:2022xfd}. This is a first step towards constructing effective theories which include fluctuations of the amplitude of the order parameter apart from its phase.

The hydrodynamic limit of hologrpahic superfluids has been studied extensively since their first discovery. As standard in hydrodynamics, the stress tensor of the theory and the electric current admit a derivative expansion in terms of the local temperature, the normal fluid velocity, the chemical potential and the phase of the order parameter. The inequivalent terms that can appear in the first few orders of the hydrodynamic series of relativistic superfluids have been classified in \cite{Herzog:2011ec,Bhattacharya:2011tra,Bhattacharya:2011eea}. This is an necessary step in order to extract the number of the transport coefficients that parametrise the different terms in the expansion series. These numbers are essentially the invariants that one can have under different choices of a fluid frame. For an isotropic relativistic fluid, the coefficients that need to be specified in the small superfluid velocity limit are the incoherent electric conductivity $\sigma$, the shear viscosity $\eta$ and three bulk viscosities $\zeta_i$. Similarly to normal fluids, conformal symmetry constrains the form of the bulk viscosities allowing only one of them to be non-zero \cite{Herzog:2011ec,Bhattacharya:2011tra}. By introducing scales through relevant scalar operators, we will retain as many independent transport coefficients as in any  relativistic superfluid.

A number of previous works have considered various aspects of the hydrodynamic limit of holographic superfluids. However, most of them have either resorted to numerical techniques \cite{Amado:2009ts,Amado:2013xya,Bhaseen:2012gg,Arean:2021tks} or they have focused on specific models where analytic solutions for the gravitational problem can be obtained infinitesimally close to the transition \cite{Herzog:2010vz,Herzog:2011ec}. In this paper, we will employ the techniques developed in \cite{Donos:2021pkk,Donos:2022xfd}, based on the Crnkovic-Witten symplectic current \cite{Crnkovic:1986ex}, to derive the first dissipative corrections in the hydrodynamic limit of holographic superfluids. A significant advantage of this approach is that an explicit solution of the gravitational fields is not required. Instead, it focuses on the universal aspects of the black holes dual to the thermal states.

As a by-product of our derivation, we will be able to fix the five non-trivial dissipative transport coefficients that we expect to have. The expressions we will obtain will be in terms of horizon data of the black holes dual to the thermal states of our system. Our results reproduce a well known result for the shear viscosity $\eta$ to entropy density $s$ ratio which is proportional to $1/4\pi$ \cite{Son:2007vk,Iqbal:2008by,Herzog:2011ec}. However, our results for the incoherent conductivity and the bulk viscosities are new. In particular, our formula for the bulk viscosity $\zeta_1$ generalises the results of \cite{Eling:2011ms} and \cite{Donos:2022uea} to include the bulk massive vector on the horizon. Moreover, as we will see our expression for $\zeta_3$ generalises the result of \cite{Donos:2021pkk} which was obtained for holographic superfluids at zero chemical potential.

Given our analytic expressions in terms of horizon data, we are able study the behaviour of the first dissipative corrections close to the phase transition. By following general arguments about the behaviour of our bulk fields near the phase transition, we are able to show that the shear viscosity and the electric conductivity are continuous functions across the transition. At the same time, we show that all three bulk viscosities diverge at the critical point.

As one might anticipate, the hydrodynamic modes of our system consist of the two longitudinal sound modes and the transverse shear mode responsible for the diffusion of momentum density. Taking the limit of the dispersion relations close to the critical temperature we find that the speed of the first sound mode remains finite while the speed for the second sound vanishes. These results can be shown through general considerations of ideal superfluid hydrodynamics. More interestingly, using the explicit expressions for the dissipative coefficients we show the the attenuation of the first sound diverges. This is in contrast to the second sound whose attenuation part remains finite.

Our paper is organised in six sections which are further divided in subsections. In section \ref{sec:setup} we present the class of holographic models we wish to study along with the thermodynamic properties of the geometries dual to the field theory thermal states. In section \ref{sec:pert} we discuss perturbations around our black holes and present the static perturbations which are the infinite wavelength and zero frequency limits of our hydrodynamic expansion. In section \ref{sec:const_rel} we extract the leading dissipative corrections to the ideal superfluid based on the techniques we developed in \cite{Donos:2021pkk,Donos:2022xfd}. We conclude our analytic results in section \ref{sec:limits} where we consider the limit of our hydrodynamic expansion close to the critical point. We also take the limit of zero chemical potential and compare our results with those in \cite{Donos:2021pkk}.

\section{Setup}\label{sec:setup}
To model a holographic superfluid phase at finite density, we will consider a bulk theory which contains a Maxwell field $A_{\mu}$, a neutral scalar $\phi$ and a complex scalar $\psi$ which is charged under the local $U(1)$ symmetry. The neutral scalar $\phi$ is not a necessary ingredient but we will use it in order to introduce additional scales into the system. This will allow all the bulk viscosities we expect to find to be non-zero.

The system is described by the bulk action,
\begin{align}\label{eq:bulk_action}
S=\int d^{4}x\,\sqrt{-g}\,\left(R-V(\phi,|\psi|^{2})-\frac{1}{2}\partial_{\mu}\phi\,\partial^{\mu}\phi-\frac{1}{2}(D_{\mu}\psi)(D^{\mu}\psi)^{\ast}-\frac{1}{4}\tau(\phi,|\psi|^{2})\,F^{\mu\nu}F_{\mu\nu} \right)\,,
\end{align}
with the covariant derivative $D_{\mu}\psi=\nabla_{\mu}\psi+iq_e A_{\mu}\,\psi$ and the field strength $F=dA$. It is easy to see that the above action is invariant under the local gauge transformations $A\to A+d\Lambda$ and $\psi\to e^{-iq_e \Lambda}\,\psi$.

Our focus will be on the superfluid phase of our system corresponding to backgrounds with a non-trivial profile for the complex bulk scalar $\psi$. In this case, the field redefinition $\psi=\rho\,e^{iq_e\theta}$ is well defined bringing the action to the form,
\begin{align}
S=\int d^{4}x\,\sqrt{-g}\,\left(R-V(\phi,\rho^{2})-\frac{1}{2}(\partial\phi)^{2}-\frac{1}{2}(\partial\rho)^{2}-\frac{1}{2}q_e^2\,\rho^{2}\,B^{2}-\frac{1}{4}\tau(\phi,\rho^{2})\,F^{\mu\nu}F_{\mu\nu} \right)\,,
\end{align}
where we have set $B=A+\partial\theta$ and $F=dB$. The equations of motion which extremise the bulk action are,
\begin{align}\label{eq:eom}
R_{\mu\nu}-\frac{1}{2}g_{\mu\nu}V-\frac{\tau}{2}\left( F_{\mu\rho}F_{\nu}{}^{\rho}-\frac{1}{4}g_{\mu\nu}\,F^{2}\right) \qquad\qquad\qquad\qquad\qquad\qquad&\nn
-\frac{1}{2}\partial_{\mu}\phi\,\partial_{\nu}\phi-\frac{1}{2}\partial_{\mu}\rho\,\partial_{\nu}\rho-\frac{1}{2}q_e^2 \rho^{2}B_\mu\,B_\nu=&0\,,\nn
\nabla_{\mu}\nabla^{\mu}\phi-\partial_{\phi}V-\frac{1}{4}\partial_\phi\tau\,F^{2}=&0\,,\nn
\nabla_{\mu}\nabla^{\mu}\rho-\partial_{\rho^{2}}V\,\rho-\frac{1}{4}\partial_{\rho^{2}}\tau\,\rho\,F^{2}-q_e^2\,\rho\,B^{2}=&0\,,\nn
\nabla_{\mu}\left(\rho^{2}B^\mu\right)=&0\,,\nn
\nabla_{\mu}(\tau\,F^{\mu\nu})-q_e^2\,\rho^{2}\,B^\nu=&0\,.
\end{align}
We will consider electrically charged black brane solutions which are dual to thermal states of the deformed CFT by the relevant operator $\mathcal{O}_{\phi}$ dual to the bulk field $\phi$. Moreover, we will assume that below a critical temperature $T_{c}$, the system exhibits spontaneous breaking of the field theory global $U(1)$.

The corresponding ansatz for the background fields is,
\begin{align}\label{eq:background}
ds^{2}&=-U(r)\,dt^{2}+\frac{dr^{2}}{U(r)}+e^{2g(r)}\,\left(dx^{2}+dy^{2} \right)\,,\nn
B=&a(r)\,dt,\qquad \phi=\phi(r),\qquad \rho=\rho(r)\,.
\end{align}
The above choice of coordinates fixes the radial coordinate apart from a global shift. We will use this freedom to always have the event horizon sitting at $r=0$. Near the horizon, regularity implies the Taylor expansions
\begin{align}\label{eq:nh_bexp}
&U(r)\approx\,4\pi T\,r+\mathcal{O}(r^{2}),\qquad g(r)\approx g^{(0)}+\mathcal{O}(r),\qquad a(r)\approx r\,a^{(0)}+\cdots \,,\nn
&\phi(r)\approx \phi^{(0)}+\mathcal{O}(r),\qquad \rho(r)\approx \rho^{(0)}+\mathcal{O}(r)\,,
\end{align}
where $T$ is the Hawking temperature.

Close to the conformal boundary at $r\to\infty$ we impose the expansions,
\begin{align}
&U(r)\approx (r+R)^{2}+\cdots+\frac{g_{(v)}}{r+R}+\cdots,\qquad g(r)\approx \ln(r+R)+\mathcal{O}(r^{-1})\,,\nn
&a(r)\approx \mu_t-\frac{\varrho}{r+R}+\cdots\,,\nn
&\phi(r)\approx \phi_{(s)}\,(r+R)^{\Delta_{\phi}-3}+\cdots +\phi_{(v)}\,(r+R)^{-\Delta_{\phi}}+\cdots\,,\nn
&\rho(r)\approx \rho_{(s)}\,(r+R)^{\Delta_{\psi}-3}+\cdots  \rho_{(v)}\,(r+R)^{-\Delta_{\psi}}+\cdots\,,
\end{align}
where we have defined the scalar operators sources $\phi_{(s)}$ and $\rho_{(s)}$ for the neutral and complex operators correspondingly. Moreover, we have defined the chemical potential $\mu_t$ and as we will see in the next section, the field theory charge density is given by the constant of integration $\varrho$. In this paper we will consider spontaneous breaking of the global $U(1)$ and we will be setting the non-perturbative complex scalar source $\rho_{(s)}$ equal to zero. Finally, the global shift in the radial coordinate which fixes the horizon at $r=0$ is reflected by the constant of integration $R$.

In general, when the complex scalar source $\rho_{(s)}$ is set to zero, a perturbation of the bulk vector will admit the UV expansion \cite{Donos:2021pkk},
\begin{align}\label{eq:uv_bexp}
\delta B_\alpha=\frac{\partial_\alpha \delta\theta_{(s)}}{(r+R)^{3-2\,\Delta_\rho}}+\cdots+ \delta m_\alpha +\cdots+\frac{ \delta j_\alpha}{r+R}+\cdots\,,
\end{align}
where $m_{\alpha}=\partial_{\alpha}\theta_{(v)}+\mu_{\alpha}$ is a gauge invariant combination of the superfluid velocity $\partial_{\alpha}\theta_{(v)}$ and the source $\mu_{\alpha}$ for the $U(1)$ current. As we will explain in the next section, the constant of integration $\theta_{(s)}$ is essentially a source for the complex scalar operator and the constants $\delta j_\alpha$ correspond to perturbations of the $U(1)$ electric current. As we will see these in the next section, these constants of integration are subject to a scalar constraint which is equivalent to the Ward identity satisfied by the electric current.

From the above we see that the phase $\theta$ of the complex scalar $\psi$ has become part of the massive vector $B$. It will useful for us to note that, in the absence of a background source $\rho_{(s)}$ the asymptotic expansion for the bulk phase close to the conformal boundary is given by,
\begin{align}
\delta\theta\approx (r+R)^{2\Delta_\psi-3}\,\delta\theta_{(s)}+\cdots+\delta\theta_{(v)}+\cdots\,,
\end{align}
allowing us to write the expansion for the complex scalar perturbation,
\begin{align}
\psi&\approx (r+R)^{\Delta_\psi-3}\,e^{iq_e\,\theta_{(v)}}\,(iq_e\,\rho_{(v)}\,\delta\theta_{(s)}+\delta\rho_{(s)})\nn
&\qquad\qquad+\cdots+(r+R)^{-\Delta_\psi}\,e^{iq_e\,\theta_{(v)}}\,(\delta\rho_{(v)}+iq_e\,\rho_{(v)}\,\delta\theta_{(v)})+\cdots\,.
\end{align}
From the above, we can read off the perturbative source for the complex operator to be,
\begin{align}\label{eq:uv_thetaexp}
\delta\lambda=e^{iq_e\,\theta_{(v)}}\,(iq_e\,\rho_{(v)}\,\delta\theta_{(s)}+\delta\rho_{(s)})\,.
\end{align}

\subsection{Thermodynamics and Ward Identities}
In order to extract quantities which are relevant to the field theory living on the boundary, our bulk action \eqref{eq:bulk_action} needs to be supplemented with appropriate counterterms that will render it finite on-shell \cite{Skenderis:2002wp}. Apart from regularisation, the appropriate counterterms make the variational problem well defined \cite{Papadimitriou:2005ii}. For the bulk action we are interested in, a set universal terms are contained in,
\begin{align}\label{eq:bdy_action}
S_{bdr}=&-\int_{\partial M}d^{3}x\,\sqrt{-\gamma}\,\left(-2K + 4 +R_{bdr}\right)\notag\\
&\quad -\frac{1}{2}\int_{\partial M}d^{3}x\,\sqrt{-\gamma}\,[(3-\Delta_{\phi})\phi^2-\frac{1}{2\Delta_{\phi}-5}\,\partial_{a}\phi\partial^{a}\phi]\notag\\
&\quad -\frac{1}{2}\int_{\partial M}d^{3}x\,\sqrt{-\gamma}\,[(3-\Delta_{\psi})|\psi|^2-\frac{1}{2\Delta_{\psi}-5}\,D_{a}\psi D^{a}\psi^{\ast}]+\cdots\,,
\end{align}
where $\gamma_{\alpha\beta}$ is the induced metric on the asymptotic hypersusface $\partial M$ of constant radial coordinate $r$.

The grand canonical free energy density $w_{FE}$ for the thermal states captured by the geometries of equation \eqref{eq:background} coincides with the value of the total action $S_{tot}=S+S_{bdr}$ after a Wick rotation to Euclidean time $t=-i\,\tau$. This yields the total Euclidean action $I_{tot}$ such that $w_{FE}=T\,I_{tot}$ and,
\begin{align}\label{eq:w_fe_def}
w_{FE}=\epsilon-T\,s-\mu_t\,\varrho\,.
\end{align}
In the expression above, $\epsilon$ is the conserved energy density, $\varrho$ is the electric charge density and,
\begin{align}\label{eq:entropy_density}
s=4\pi\,e^{2\,g^{(0)}}\,,
\end{align}
is the Bekenstein-Hawking entropy density of the system. Another horizon quantity we can define and which will become useful later is related to the flux of the one-from $B_\mu$ through the black hole horizon. We will can this the horizon charge density $\varrho_h$ and by using our near horizon expansion \eqref{eq:nh_bexp} we can write,
\begin{align}\label{eq:hcharge_density}
\varrho_h=e^{2g^{(0)}}\tau^{(0)}a^{(0)}\,.
\end{align}

In our analysis, we are aiming to use the set of techniques that were developed in \cite{Donos:2021pkk,Donos:2022xfd,Donos:2022uea} in order to extract holographic information for the perturbations in real time. For this reason, we will perform an integration by parts in the Einstein-Hilbert in order to obtain a Lagrangian density that will only contain first order partial derivatives of the metric. As we argued in \cite{Donos:2022uea}, the boundary counterterms $\delta S^\prime_{bdr}$ for that action will be given by,
\begin{align}
\delta S^\prime_{bdr}=&-\int_{\partial M}d^{3}x\,\sqrt{-\gamma}\,\left(4 +R_{bdr}\right)\notag\\
&\quad -\frac{1}{2}\int_{\partial M}d^{3}x\,\sqrt{-\gamma}\,[(3-\Delta_{\phi})\phi^2-\frac{1}{2\Delta_{\phi}-5}\,\partial_{a}\phi\partial^{a}\phi]\notag\\
&\quad -\frac{1}{2}\int_{\partial M}d^{3}x\,\sqrt{-\gamma}\,[(3-\Delta_{\psi})|\psi|^2-\frac{1}{2\Delta_{\psi}-5}\,D_{a}\psi D^{a}\psi^{\ast}]+\cdots\,.
\end{align}
After integrating by parts, the total action is given by,
\begin{align}\label{eq:first_order_action}
S^\prime=\int\,d^4x\, \mathcal{L}\left(g_{\mu\nu},\partial_\lambda g_{\mu\nu},\varphi^I,\partial_\lambda\varphi^I\right)+S_{bdr}^\prime\,.
\end{align}
As we explained in \cite{Donos:2022uea}, this action is suitable to extract real time boundary quantities as we drop a term that can potentially arise on the black holes horizon. Notice that dropping this term is standard within the holographic dictionary. However, for the Euclidean solutions this term is responsible for the entropy term in the expression \eqref{eq:w_fe_def} for the grand canonical free energy.

The VEV of the boundary stress tensor $T_{\mu\nu}$, the electric current $J_\mu$ and the scalar operators $\mathcal{O}_\phi$ and $\mathcal{O}_\psi$ are obtained via the variations,
\begin{align}
\langle T^{\mu\nu}\rangle&=\lim_{r\to\infty}\frac{2\,r^5}{\sqrt{-\gamma}}\left[ \frac{\partial\mathcal{L}}{\partial(\partial_r g_{\mu\nu})}+\frac{\delta S_{bdr}^\prime}{\delta\gamma_{\mu\nu}}\right]\,,\nn
\langle J^{\mu}\rangle&=\lim_{r\to\infty}\frac{r^5}{\sqrt{-\gamma}}\left[ \frac{\partial\mathcal{L}}{\partial(\partial_r B_{\mu})}+\frac{\delta S_{bdr}^\prime}{\delta B_{\mu}}\right]\,,\nn
\langle \mathcal{O}_\phi\rangle&=\lim_{r\to\infty} \frac{r^3}{\sqrt{-\gamma}}\left[\frac{\partial\mathcal{L}}{\partial(\partial_r \phi)}+ \frac{\delta S_{bdr}^\prime}{\delta \phi}\right]\,,\nn
\langle \mathcal{O}_\psi\rangle&=\lim_{r\to\infty} \frac{r^3}{\sqrt{-\gamma}}\left[\frac{\partial\mathcal{L}}{\partial(\partial_r \psi^\ast)}+ \frac{\delta S_{bdr}^\prime}{\delta \psi^\ast}\right]\,,
\end{align}
This form will be particularly useful to us as we will be able to read off directly the dissipative parts of the VEVs for the stress tensor and the electric current by using the techniques developed in \cite{Donos:2021pkk,Donos:2022xfd,Donos:2022uea}.

In our analysis, the gravitational and vector field constraints in the bulk will be the last set of equations to be imposed. From the field theory point of view, when imposed on a hypersurface close to conformal boundary, they are equivalent to the Ward identities of diffeomorphism and gauge invariance for the sources,
\begin{align}\label{eq:Ward}
\nabla_a \langle T^{ab}\rangle&=F^{ba}\langle J_a \rangle+\nabla^{b}\varphi_{(s)}^I\,\langle \mathcal{O}_I\rangle+\nabla^{b}\lambda\,\langle \mathcal{O}_{\psi^\ast}\rangle+\nabla^{b}\lambda^\ast\,\langle \mathcal{O}_{\psi}\rangle\,,\nn
\nabla_a\langle J^a\rangle&=\frac{q_e}{2 i}\,\left(\lambda\,\langle \mathcal{O}_{\psi^\ast}\rangle-\lambda^\ast\,\langle \mathcal{O}_{\psi}\rangle \right)\,.
\end{align}
In the above we have define the field strength $F=dm=d\mu$ of the external source of the current operator. Our main aim is to first obtain a set of constitutive relations for the stress tensor and electric current in terms of an appropriate set of hydrodynamic variables by solving the radial equations. The Ward identities \eqref{eq:Ward} will play the role of conservation laws in the theory of hydrodynamics.

The background geometries of equation \eqref{eq:background} yield the stress tensor,
\begin{align}\label{eq:bac_stress}
\langle T_{tt}\rangle=\epsilon\,,\quad \langle T_{xx} \rangle= \langle T_{yy} \rangle=p\,,\quad \langle J^{t}\rangle=\varrho\,,
\end{align}
where the energy density $\epsilon$ is given by,
\begin{align}
\epsilon=-2\,g_{(v)}-\varphi_{(s)}\,\langle \mathcal{O}_\phi\rangle\,,
\end{align}
with $\langle\mathcal{O}_\phi\rangle=(2\Delta_\phi-3)\,\phi_{(v)}$. Moreover, due to the translational invariance of the background thermal states under consideration, the pressure $p$ is related to the free energy density according to $p=-w_{FE}$. 

Finally, the VEV of our complex scalar operators is given by the constants of integration appearing in the asymptotic expansions \eqref{eq:uv_bexp} and \eqref{eq:uv_thetaexp} according to,
\begin{align}
\langle\mathcal{O}_\psi\rangle=\frac{1}{2}(2\Delta_\psi -3)\,\rho_{(v)}e^{i\,q_e\,\theta_{(v)}}\,,
\end{align}
which after we perturb yields,
\begin{align}\label{eq:o_psi_var}
\delta\langle\mathcal{O}_\psi\rangle=\langle\mathcal{O}_\psi\rangle_b\,i\,q_e\,\delta\theta_{(v)}+e^{i\,\arg(\langle\mathcal{O}_\psi\rangle_b)}\, \delta|\langle\mathcal{O}_\psi\rangle|\,.
\end{align}

The thermal states we are considering in equation \eqref{eq:background} are parametrised by temperature $T$, the chemical potential $\mu_t$ and the neutral scalar deformation parameter $\phi_{(s)}$. However, superfluids are also characterised by a non-trivial electric current susceptibility $\chi_{JJ}$ allowing for thermal states which have a persistent electric with no heat current flow. This leads to a form of the first law,
\begin{align}
dw_{FE}=-s\,dT-j^a\,d m_a-\langle\mathcal{O}_\phi\rangle\,d\phi_{(s)}\,,
\end{align}
where once again we have the gauge invariant combination $m_a=\mu_a+\partial_a\theta_{(v)}$. Even though we are not considering thermal background states with a persistent current, their perturbative form will become important in the construction of our hydrodynamic modes, through the supercurrent susceptibility. We will come back to this point in the next subsection when we discuss our static, thermodynamic perturbations.

For later reference, we will also define the thermodynamic susceptibilities through the variations,
\begin{align}\label{eq:therm_susc}
ds=T^{-1}\,c_\mu\,dT+\xi\,d\mu_t\,,\quad\quad d\varrho=\xi\,dT+\chi_{QQ}\,d\mu_t\,.
\end{align}

\section{Perturbations}\label{sec:pert}

In this section we will discuss the various perturbations around the black hole geometries \eqref{eq:background} that will be relevant to our construction. In subsection \ref{sec:pert_realt} we discuss properties of general space-time dependent perturbations before considering any hydrodynamic limit. In subsections \ref{sec:pert_thermo} and \ref{sec:pert_diffeo} we discuss static perturbations that we can obtain through variations of the thermodynamic variables parametrising our background geometries as well as perturbations that we can  obtain through large coordinate transformations. As we will see, these will play a dual role in obtaining the constitutive relations. Finally, in subsection \ref{sec:hydro} we present our hydrodynamic expansion along with the leading dissipative corrections we are after. 

\subsection{Real Time Perturbations}\label{sec:pert_realt}

Before discussing the derivative expansion of hydrodynamics, it will be beneficial to consider space-time dependent fluctuations from the boundary theory point of view. Due to the translational symmetry in both space and time, we find it convinient to perform a Fourier mode expansion of the general form,
\begin{align}\label{eq:fourier_modes}
\delta\mathcal{F}(t,x^i;r)=e^{-iw\,(t+S(r))+ik_i x^i}\,\delta f(r)\,,
\end{align}
where $\delta\mathcal{F}$ represents represents perturbations of the scalars as well as the metric field components. Moreover, by choosing the function $S(r)$ to approach $S(r)\to\frac{\ln r}{4\pi T}+\cdots$ close to the black hole horizon at $r=0$, we are guaranteed to the correct infalling boundary conditions provided that $\delta f(r)$ admits a Taylor series expansion there. Finally, in order to for the holographic dictionary to be solely dictated by the asymptotics of $\delta f(r)$, we will choose $S(r)$ to behave as $S(r)\to \mathcal{O}(1/r^3)$ close to the conformal boundary.

Close to the conformal the radial functions that parametrise our space-time dependent perturbations will have to behave according to,
\begin{align}\label{eq:pert_uv_bcs}
\delta g_{ab}(r)&=(r+R)^2\,\left(\delta s_{ab}+\cdots +\frac{\delta t_{ab}}{(r+R)^3}+\cdots\right)\,,\nn
\delta g_{ra}(r)&=\mathcal{O}\left(\frac{1}{r^3}\right),\quad\delta g_{rr}(r)=\mathcal{O}\left(\frac{1}{r^4} \right)\,,\nn
\delta B_a(r)&=\delta m_a+\frac{\delta j_a}{r+R}+\cdots,\quad \delta B_r(r)=\mathcal{O}\left( \frac{1}{r^3}\right)\,,\nn
\delta\phi(r)&=\frac{\delta\phi_{(v)}}{(r+R)^{\Delta_\phi}}+\cdots,\quad \delta\rho(r)=\frac{\delta\rho_{(v)}}{(r+R)^{\Delta_\psi}}+\cdots\,.
\end{align}
Notice that we do not choose to work in a particular coordinate system, we will only our coordinates asymptotically through the decays of the metric components $\delta g_{r\mu}$. Moreover, the above stated decay for the one-form field components $\delta B_r$ is correct provided that we impose the Ward identity for the electric current in equation \eqref{eq:Ward}.

Close to the black hole horizon at $r=0$, we need to impose ingoing boundary conditions which we can achieve through the asymptotics,
\begin{align}\label{eq:gen_exp}
\delta g_{tt}(r)&= 4\pi T\,r\, \delta g_{tt}^{(0)}+\cdots\,,\quad
\delta g_{rr}(r)=\frac{\delta g_{rr}^{(0)}}{4\pi T\,r}+\cdots\, \,,\nn
\delta g_{ti}(r)&=\delta g_{ti}^{(0)}+r\,\delta g_{ti}^{(1)}+\cdots\,,\quad 
\delta g_{ri}(r)=\frac{\delta g_{ri}^{(0)}}{4\pi T\,r}+\delta g_{ri}^{(1)}+\cdots\,,\nn
\delta g_{ij}(r)&=\delta g_{ij}^{(0)}+\cdots\,,\quad\quad
\delta g_{tr}(r)=\delta g_{tr}^{(0)}+\cdots\,,\nn
\delta B_t(r)&=\delta b_t^{(0)}+\delta b_t^{(1)}\,r+\cdots,\quad\quad \delta B_r(r)=\frac{\delta b_r^{(0)}}{4\pi T\,r}+\delta b_r^{(1)}+\cdots\,,\nn
\delta B_i(r)&=\delta b_i^{(0)}+\cdots\,,\nn
\delta \phi(r)&=\delta \phi^{(0)}+\cdots,\quad\quad \delta \rho(r)=\delta \rho^{(0)}+\cdots\,.
\end{align}
In order to achieve regular infalling boundary conditions, the above need to be supplemented by additional conditions,
\begin{align}\label{eq:nh_reg}
-2\pi T(\delta g_{tt}^{(0)}+\delta g_{rr}^{(0)})=-4\pi T\,\delta g_{rt}^{(0)}&\equiv \delta T_h\,,\notag\\
\delta g_{ti}^{(0)}=\delta g_{ri}^{(0)}&\equiv-\delta u_{i}\,,\nn
\delta b_r^{(0)}=\delta b_t^{(0)}&\equiv \delta\mu_h\,.
\end{align}
In the above equations we have defined some horizon quantities which can be thought of as local temperature, fluid velocity and chemical potential and which can be defined on the black hole horizon. These will not be directly relevant to us since we will study the fluid from the boundary theory point of view.

\subsection{Thermodynamics Perturbations}\label{sec:pert_thermo}
In this subsection we will consider all static perturbations which can be realised as small variations of thermodynamic variables and the Goldstone mode. As such, these will be the starting point of our hydrodynamic expansion which we will have to correct by using the techniques introduced in \cite{Donos:2021pkk,Donos:2022xfd,Donos:2022uea}.

The first perturbation we will consider is generated by a variation in the temperature $T$ of the background black holes of equation \eqref{eq:background}. However, in order to achieve the regular infalling boundary conditions of equations \eqref{eq:gen_exp} and \eqref{eq:nh_reg}, we need to accompany the temperature variation with an infinitesimal coordinate transformation,
\begin{align}
t\to t-\partial_T S\,\delta T\,,
\end{align}
yielding the fluctuation, 
\begin{align}\label{eq:static_pert_dT}
\delta_T g_{tt}&=-\partial_T U\,,\quad \delta_T g_{rr}=-\frac{\partial_T U}{U^2}\,,\quad \delta_T g_{tr}=U\,\partial_T S^\prime\,,\nn
\delta_T g_{ij}&=2\,\delta_{ij}\,e^{2g}\,\partial_T g\,,\quad \delta_T B_t=\partial_T a\,,\nn
\delta_T\phi&=\partial_T\phi\,,\quad \delta_T\rho=\partial_T\rho\,.
\end{align}

Another thermodynamic variation which we wish to consider is with respect to the external field source $\delta\mu_a$ and superfluid velocity $\partial_a\delta\theta$. These two variables are packaged in the gauge invariant quantity $m_a=\mu_a+\partial_a\theta$. In order to generate the variations with respect to $m_t$, we can simply consider the derivative of our backgrounds with respect to the chemical potential to obtain,
\begin{align}\label{eq:static_pert_dmut}
\delta_{m_t} g_{tt}&=-\partial_{\mu_t} U\,,\quad \delta_{m_t} g_{rr}=-\frac{\partial_{\mu_t} U}{U^2}\,,\quad \delta_{m_t} g_{ij}=2\,\delta_{ij}\,e^{2g}\,\partial_{\mu_t} g\nn
\delta_{m_t} B_t&=\partial_{\mu_t} a\,,\quad \delta_{m_t}\phi=\partial_{\mu_t}\phi\,,\quad \delta_{m_t}\rho=\partial_{\mu_t}\rho\,.
\end{align}
In contrast to the case of the temperature variation, we don't need to shift the time coordinate here. The derivative of $U$ with respect to the chemical potential results in a perturbation such that $\delta g_{tt}$ and $\delta g_{rr}$ exhibit a regular behaviour near the horizon, as can be seen from the asymptotics \eqref{eq:nh_bexp}. In the notation of equations \eqref{eq:gen_exp} and \eqref{eq:nh_reg} it leads to $\delta T_{h}=0$.

The perturbation with respect to the spatial components $m_i$ will result in a static solution representing a static, isotropic current flow captured by,
\begin{align} \label{eq:static_pert_dmui}
\delta_{m_j} g_{ti}&=\delta^j_i\,\delta f_g\,,\quad \delta_{m_j}B_i=\delta^j_i\,\delta f_b\,.
\end{align}
Since this solution is not directly generated through a derivative of our backgrounds \eqref{eq:background}, it is worth discussing it in more detail. An important point is that in order to define the current-current susceptibility $\chi_{JJ}$ through this fluctuation, we have to fix a frame where there is no heat current flow. This is necessary in order to obtain a unique solution since in the opposite case, we would be able to perform a Lorentz boost an still have a solution with a different current and heat flow. From a physics point of view, this is simply the fact that we want to consider a frame where only the superfluid component carries electric current. In this situation, there cannot be a heat current as the superfluid cannot transport it.

The absence of a heat current flow on the boundary imposes that the constant term in $\delta f_g(r)$ will vanish in the near horizon expansion,
\begin{align}
\delta f_g(r)=\delta f_g^{(1)}\,r+\cdots\,,\quad \delta f_b(r)=\delta f_b^{(0)}+\cdots\,.
\end{align}
This can be justified through either earlier works relating the horizon data of static perturbations to boundary quantities \cite{Donos:2014cya,Donos:2015gia} or by following the techniques used in the present paper. By using our techniques, we can show this statement by e.g. considering the symplectic current constructed from the perturbation \eqref{eq:static_pert_dmui} and $\delta_{s_{ti}}$ discussed in the next subsection.  This is also the reason that we don't need to include a perturbation for the metric component $\delta g_{ri}$ since this produces a perturbation which has $\delta u_{i}=0$, in the notation of the boundary conditions \eqref{eq:gen_exp} and \eqref{eq:nh_reg}. Asymptotically we must have,
\begin{align}
\delta f_g(r)=\frac{1}{3}\frac{\mu_t\,\chi_{JJ}}{r+R}+\cdots\,,\quad \delta f_b=1-\frac{\chi_{JJ}}{r+R}+\cdots\,,
\end{align}
which imposes the zero heat current condition.

The final static solution we would like to consider is generated by boundary by infinitesimal Lorentz boosts with parameter $\delta v_i$. As one can easily see, at finite chemical potential this can be generated by a combination of a metric and a constant current source,
\begin{align}
t\to t- \delta v_i\,x^i\,,\quad x^i\to x^i-\delta v^i\,t\,,\quad \delta m_i=\mu_t\,\delta v_i\,,
\end{align}
which will guarantee the absence of a net current source in the boundary theory. In order to obtain a regular solution in the bulk, we need to see the above as the asymptotics of a large coordinate transformation which is otherwise regular everywhere in the bulk. This is achieved by the coordinate transformation,
\begin{align}
t\to t- \delta v_i\,x^i\,,\quad x^i\to x^i-\delta v^i\,(t+S(r))\,,
\end{align}
combined with the static solution \eqref{eq:static_pert_dmui} with the appropriate sources to cancel the external gauge field source. This leads to the bulk perturbation,
\begin{align} \label{eq:static_pert_dv}
\delta_{v_j} g_{tt}&=\delta_{v_j} g_{rr}=\delta_{v_j} g_{tr}=0\,,\quad \delta_{v_j}\varphi=0\,,\nn
\delta_{v_j} g_{ti}&=\delta^j_i\,\left(U-e^{2g}+\mu_t\,\delta f_g\right)\,,\quad \delta_{v_j} g_{ri}=-\delta^j_i\,\left(e^{2g}\,S^\prime +\mu_t\,\delta f_c\right)\,,\nn
\delta_{v_j}B_i&=\delta^j_i\,\left(\mu_t\,\delta f_b-a \right) \,,\quad \delta_{v_j}B_t=0\,,\quad \delta_{v_j}B_r=0\,.
\end{align}

The perturbations discussed in this section will be the building blocks for the zero frequency, infinite wavelength limit of the hydrodynamic perturbations we will start constructing in section \ref{sec:hydro}. From the boundary point of view, the perturbations we discussed here give rise to the stress tensor and electric current fluctuations of the ideal superfluid,
\begin{align}\label{eq:VEVs_pert_thermo}
\delta\langle T^{tt}\rangle&=\partial_T\epsilon\,\delta T+\partial_{\mu_t}\epsilon\,\delta m_t=\left(c_{\mu}+\mu_t\,\xi\right)\,\delta T+\left(T\,\xi+\mu_t\,\chi_{QQ} \right)\,\delta m_t\,,\nn
\delta\langle T^{ij}\rangle&=\delta^{ij}\,\left(\partial_T p\,\delta T+\partial_{\mu_t}p\,\delta m_t \right)=\delta^{ij}\,\left(s\,\delta T+\varrho\,\delta m_t \right)\,,\nn
\delta\langle T^{ti}\rangle&=\left( \epsilon+p-\mu_t^2\,\chi_{JJ}\right)\,\delta v^i-\mu_t\,\chi_{JJ}\,\delta m^i\,,\nn
\delta\langle J^t\rangle&=\partial_T\varrho\,\delta T+\partial_{\mu_t}\varrho\,\delta m_t=\xi\,\delta T+\chi_{QQ}\,\delta m_t\,,\nn
\delta\langle J^i\rangle&=\left( \varrho-\mu_t\,\chi_{JJ}\right)\,\delta v^i-\chi_{JJ}\,\delta m^i\,.
\end{align}
It is also useful to check the heat current,
\begin{align}
\delta\langle Q^i\rangle=\delta \langle T^{ti}\rangle-\mu_t\,\delta\langle J^i\rangle=\left( \epsilon+p-\mu_t\,\varrho\right)\,\delta v^i=T s\,\delta v^i\,,
\end{align}
is indeed transferred by the normal fluid component which contributes to the entropy density of the system. Moreover, it will be useful to define the normal and superfluid charge densities,
\begin{align}\label{eq:rsrn}
\varrho_n&=\varrho-\mu_t\,\chi_{JJ}\,,\nn
\varrho_s&=\mu_t\,\chi_{JJ}\,,
\end{align}
respectively.

\subsection{Static Perutrbations from Diffeomorphisms}\label{sec:pert_diffeo}
Another class of static perturbations we will find useful are simply generated by large diffeomorphisms. Similarly with the thermodynamic perturbations of the previous subsection, their role in our construction will be twofold. Firstly, they will later be used in order to introduce sources in our hydrodynamic expansion as they will be part of the zero frequency, infinite wavelength limit. Secondly, they will be used in the construction of the symplectic current. As described in \cite{Donos:2022uea}, this will help us read off the leading dissipative corrections to the stess tensor.

Similarly to \cite{Donos:2022uea}, we will consider the combination of large coordinate transformations along with constant sources for the electric current sources,
\begin{align}\label{eq:coord_trans_sources}
x^a&\to x^a+ \delta s^a{}_b\,(x^b+\delta^b_t\,S(r))\,,\nn
\delta\mu_a&=-\mu_t\,\delta s^t{}_a\,,
\end{align}
where the term involving $S(r)$ takes care of regular, infalling boundary conditions on the black hole horizon at $r=0$. It is easy to see that from the boundary theory point of view, this coordinate transformation induces a change in the metric,
\begin{align}
\delta g_{ab}&=\eta_{ac}\,\delta s^c{}_b+\eta_{bc}\,\delta s^c{}_a=2\,\delta s_{\left(ab\right)}\,,
\end{align}
while the explicit current sources of equation \eqref{eq:coord_trans_sources} combine with the transformation of the background source to give zero.

From the bulk point of view the bulk perturbation corresponding to the source $\delta s_{tt}$ is realised by,
\begin{align}
\delta_{s_{tt}}g_{tt}&=2\,U-\mu_t\,\partial_{\mu_t}U\,,\quad \delta_{s_{tt}}g_{tr}=U\,(S^\prime+\mu_t\,\partial_{\mu_t}S^\prime)\,,\nn
\delta_{s_{tt}}g_{rr}&=-\mu_t\,\frac{\partial_{\mu_t}U}{U^2}\,,\quad \delta_{s_{tt}}g_{ij}=2\,\delta_{ij}\,\mu_t\,e^{2g}\,\partial_{\mu_t}g\,,\nn
\delta_{s_{tt}}B_{t}&=-a+\mu_t\,\partial_{\mu_t}a\,,\quad \delta_{s_{tt}}B_{r}=-a\,S^\prime\,,\nn
\delta s_{tt}\phi&=\mu_t\,\partial_{\mu_t}\phi\,,\quad \delta s_{tt}\rho=\mu_t\,\partial_{\mu_t}\rho\,.
\end{align}
For the rest of the perturbations generated by \eqref{eq:coord_trans_sources} we have,
\begin{align}\label{eq:static_pert_source}
\delta_{s_{tj}}g_{ti}&=\delta^j_i\,\left(U+\mu_t\,\delta f_g\right)\,,\quad \delta_{s_{tj}}B_i=\delta^j_i\,\left(-a+\mu_t\,\delta f_b\right)\,,\nn
\delta_{s_{jt}}g_{ti}&=\delta^j_i\,e^{2g}\,,\quad \delta_{s_{jt}}g_{ri}=\delta^j_i\,e^{2g}\,S^\prime\,,\nn
\delta_{s_{ij}}g_{kl}&=2\,\delta^{\left(i\right.}_k \delta^{\left.j\right)}_l\,e^{2g}\,.
\end{align}

It is a simple matter to combine variations and transformations the VEVs of the background stress tensor and electric current to obtain the perturbations,
\begin{align}\label{eq:VEVs_pert_diffeo}
\delta\langle T^{tt}\rangle&=\left(2\,\epsilon+\mu_t\,\left(T\,\xi+\mu_t\,\chi_{QQ} \right)\right)\,\delta s^{tt}\,,\nn
\delta\langle T^{ij}\rangle&=\delta^{ij}\mu_t\,\varrho\,\delta s^{tt}-2\,p\,\delta s^{\left(ij\right)}\,,\nn
\delta\langle T^{ti}\rangle&=\epsilon\,\delta s^{it}-\left(p-\mu_t^2\,\chi_{JJ}\right)\,\delta s^{ti}\,,\nn
\delta\langle J^t\rangle&=\left(\varrho+\mu_t\,\chi_{QQ}\right)\,\delta s^{tt}\,,\nn
\delta \langle J^{i}\rangle&=\varrho\,\delta s^{it}+\mu_t\,\chi_{JJ}\,\delta s^{ti}\,.
\end{align}
The above variations with respect to the sources, along with the ones coming from thermodynamics in equation \eqref{eq:VEVs_pert_thermo}, will become the constitutive relations for the ideal superfluid.

\subsection{Hydrodynamic Perturbations}\label{sec:hydro}
In this subsection we will construct the hydrodynamic modes in the bulk. In order to do this, we think of them as finite frequency and long wavelength deformations of the static modes we discussed in subsections \ref{sec:pert_thermo} and \ref{sec:pert_diffeo}. To make things technically more tractable, we will consider the Fourier decomposition \eqref{eq:fourier_modes} with $k_i=\varepsilon\,q_i$ and $w=\varepsilon\,\omega$.

Our hydrodynamic can be expanded in $\varepsilon$ according to,
\begin{align}\label{eq:hydro_pert_exp}
\delta_H f&=\varepsilon\,\delta f^{(1)}+\varepsilon^2\,\delta f^{(2)}+\cdots\nn
&=\delta_T f\,\delta T+\delta_{v_i} f\,\delta v_i+\delta_{m_a} f\,\delta m_a+\delta_{s_{ab}} f\,\delta s_{ab}+\varepsilon^2\,\delta f^{(2)}+\cdots\,,
\end{align}
where $\delta f^{(1)}$ is a linear combination of the static modes that we discussed in subsections \ref{sec:pert_thermo} and \ref{sec:pert_diffeo}. After dressing this linear combination with the exponential of the Fourier modes, we expect that the radial function will admit an $\varepsilon$ expansion which after writing back in position space, we can think of as derivative corrections. In order to have this interpretation, we will need to show that the VEV corrections induced by the leading correction $\delta f^{(2)}$ can be indeed expressed in terms of derivatives of the local temperature $\delta T$, fluid velocity $\delta v_i$ and superfluid velocity $\partial_a\delta\theta$.

To organise the expansion of the boundary stress tensor and electric current, we can write,
\begin{align}\label{eq:const_rels}
\delta\langle T^{tt}\rangle&=\left(c_{\mu}+\mu_t\,\xi\right)\,\delta T+\left(T\,\xi+\mu_t\,\chi_{QQ} \right)\,\delta\mu+2\,\epsilon\,\delta s^{tt}+\varepsilon^2\,\delta\langle T^{tt}\rangle_{(2)}+\cdots\,,\nn
\delta\langle T^{it}\rangle&=\delta\langle T^{ti}\rangle=\left(Ts+\mu_t\,\varrho_n\,\right)\,\delta v^i-\varrho_s\,\delta m^i+\epsilon\,\delta s^{it}+\varepsilon^2\,\delta\langle T^{ti}\rangle_{(2)}+\cdots\,,\nn
\delta\langle T^{ij}\rangle&=\delta^{ij}\,\left(s\,\delta T+\varrho\,\delta\mu \right)-2\,p\,\delta s^{\left(ij\right)}+\varepsilon^2\,\delta\langle T^{ij}\rangle_{(2)}+\cdots\,,\nn
\delta\langle J^t\rangle&=\xi\,\delta T+\chi_{QQ}\,\delta\mu+\varrho\,\delta s^{tt}+\varepsilon^2\,\delta\langle J^t\rangle_{(2)}\,,\nn
\delta\langle J^i\rangle&=\varrho_n\,\delta v^i-\chi_{JJ}\,\delta m^i+\varrho\,\delta s^{it}+\varepsilon^2\,\delta\langle J^i\rangle_{(2)}\,,
\end{align}
where we have used our definitions \eqref{eq:rsrn} for the normal and superfluid charge densities. Moreover, we have defined the variation of the new chemical potential,
\begin{align}
\delta\mu=\delta m_t+\mu_t\,\delta s^{tt}\,,
\end{align}
which as we will later see in section \ref{sec:fluid_frames}, it will receive dissipative corrections. In the above, the corrections $\delta\langle T^{ab}\rangle_{(2)}$ and $\delta\langle J^{a}\rangle_{(2)}$ are precisely the corrections due to $\delta f^{(2)}$ in the expansion \eqref{eq:hydro_pert_exp}. Already at leading order, the $\mathcal{O}(\varepsilon)$ terms in the constitutive relations \eqref{eq:const_rels} are enough to are enough to yield the equations of motion of the ideal superfluid, when combined with the Ward identities \eqref{eq:Ward},
\begin{align}\label{eq:ideal_eoms}
&\left( Ts+\mu_t \varrho_n\right)\,\partial_t\delta v_i+\varrho_n\,\partial_t\delta m_i+s\,\partial_i\delta T-T s\,\partial_i\delta s^{tt}=0\,,\nn
&\xi\,\partial_t\delta T+\chi_{QQ}\,\partial_t\delta\mu+\varrho_n\,\partial_i\delta v^i-\chi_{JJ}\,\partial_i\delta m^i+\varrho\left(\partial_i\delta s^{it}+ \delta_{ij}\partial_t\delta s^{ij}\right)=0\,,\nn
&\left(c_{\mu}+\mu_t\,\xi\right)\,\partial_t\delta T+\left(T \xi+\mu_t \chi_{QQ} \right)\,\partial_t\delta\mu+\left( Ts+\mu_t \varrho_n\right)\,\partial_i\delta v^i\nn
&\qquad\qquad\qquad\qquad\qquad\qquad\qquad-\varrho_s\,\partial_i\delta m^i+\left( \epsilon+p\right)\left( \partial_i\delta s^{it}+\delta_{ij}\partial_t\,\delta s^{ij}\right)=0\,.
\end{align}

The task of the next section will be to employ the techniques of \cite{Donos:2022uea} and express these corrections in terms of $\delta T$, $\delta v_i$ and $\partial_a\delta\theta$. As a byproduct, we will obtain specific expressions for the shear and the three bulk viscosities, along with the two independent dissipative coefficients that enter the currents or the Josephson relation, depending on the frame one would like to use. For completeness we will express our constitutive relations in the transverse frame \cite{Herzog:2011ec,Bhattacharya:2011tra}.

\section{Constitutive Relations}\label{sec:const_rel}

In this section we will derive the constitutive relations for the stress tensor and electric current relevant to the hydrodynamic fluctuations of our system. We will achieve this in two steps which we discuss in detail in the following subsections. The first step is to extract the constitutive relations in terms of our hydrodynamic variables. As we will see, we will land in an unusual fluid frame which will include bulk integrals as part of its artefacts. In the second step we will change our description to the so called transverse frame from which we will be able to read off the shear viscosity $\eta$, the three bulk viscosities $\zeta_i$ and the incoherent conductivity $\sigma$.

\subsection{Constitutive Relations in Terms of Horizon Data}\label{sec:const_rel_horizon}
The aim of this section is to write the dissipative corrections of the constitutive relations \eqref{eq:const_rels} in terms of derivatives of the hydrodynamic variables $\delta T$, $\delta v^i$ and $\delta\theta$. In order to achieve this, we will follow closely the logic developed in \cite{Donos:2022uea}.

The main tool in our construction will be the Crnkovic-Witten symplectic current defined for any classical theory of a collection of fields $\phi^I$ whose equations of motion can be obtained from a first order Lagrangian density $\mathcal{L}(\phi^I,\partial\phi^I)$. For any two perturbations $\delta_1\phi^I$ and $\delta_1\phi^I$ around a background $\phi_b^I$ which solve the Euler-Lagrange equations of motion, the vector density,
\begin{align}\label{eq:scurrent_def}
P^\mu_{\delta_1,\delta_2}=\delta_1\phi^I\,\delta_2\left(\frac{\partial\mathcal{L}}{\partial \partial_\mu\phi^I} \right)-\delta_2\phi^I\,\delta_1\left(\frac{\partial\mathcal{L}}{\partial \partial_\mu\phi^I} \right)\,,
\end{align}
is divergence free,
\begin{align}\label{eq:div_free}
\partial_\mu P^\mu_{\delta_1,\delta_2}=0\,.
\end{align}
For completeness, we write list the contributing terms to \eqref{eq:scurrent_def},
\begin{align}\label{eq:sympl_cur_contr}
\frac{\partial\mathcal{L}}{\partial\partial_\mu g_{\alpha\beta}}&=\sqrt{-g}\,\Gamma^\mu_{\gamma\delta}\left(\,g^{\gamma\alpha}\,g^{\delta\beta}-\frac{1}{2}\,g^{\gamma\delta}g^{\alpha\beta} \right)-\sqrt{-g}\,\Gamma^\kappa_{\kappa\lambda}\,\left(g^{\mu\left(\alpha\right.}g^{\left.\beta\right)\lambda}-\frac{1}{2}\,g^{\mu \lambda}g^{\alpha\beta}\right)\,,\notag\\
\frac{\partial\mathcal{L}}{\partial\partial_\mu B_\nu}&=-\sqrt{-g}\,\tau\,F^{\mu\nu}\,,\quad \frac{\partial\mathcal{L}}{\partial\partial_\mu\phi}=-\sqrt{-g}\,\partial^\mu\phi\,,\quad \frac{\partial\mathcal{L}}{\partial\partial_\mu\rho}=-\sqrt{-g}\,\partial^\mu\rho\,,
\end{align}
through the derivatives of the first order bulk action \eqref{eq:first_order_action}.

Following very similar arguments with \cite{Donos:2022uea}, one can show that the asymptotic behaviour of the radial components of the symplectic current is,
\begin{align}\label{eq:scurrent_asymptotics}
P^r_{\delta_1,\delta_2}=&\frac{1}{r^3}\,\left(\delta_1\phi_{(s)}\,\delta_2\left(\sqrt{-\gamma}\,\langle\mathcal{O}_\phi\rangle\right)-\delta_2\varphi_{(s)}\,\delta_1\left(\sqrt{-\gamma}\,\langle\mathcal{O}_\phi\rangle\right)\right)\nn
&+\frac{1}{r^3}\,\left(\delta_1\rho_{(s)}\,\delta_2\left(\sqrt{-\gamma}\,\langle\mathcal{O}_\rho\rangle\right)-\delta_2\rho_{(s)}\,\delta_1\left(\sqrt{-\gamma}\,\langle\mathcal{O}_\rho\rangle\right)\right)\nn
&+\frac{1}{r^3}\,\left(\delta_1 m_a\,\delta_2\left(\sqrt{-\gamma}\,\langle J^a\rangle\right)-\delta_2 m_a\,\delta_1\left(\sqrt{-\gamma}\,\langle J^a\rangle\right)\right)\nn
&+\frac{1}{r^3}\frac{1}{2}\left(\delta_1\gamma_{ab}\,\delta_2\left(\sqrt{-\gamma}\,\langle T^{ab}\rangle\right)-\delta_2\gamma_{ab}\,\delta_1\left(\sqrt{-\gamma}\,\langle T^{ab}\rangle\right)\right)+\cdots\,,
\end{align}
with the VEVs being correct up to second derivatives of the sources. This is useful to us as we are interested in obtaining the VEVs of the stress tensor and the electric current up to order $\mathcal{O}(\varepsilon^2)$. The above expression shows that by using appropriate combinations of our hydrodynamic mode \eqref{eq:hydro_pert_exp} and the static solutions of subsections \ref{sec:pert_thermo} and \ref{sec:pert_diffeo}, we can read off the VEVs of our system.

By introducing the Fourier decomposition \eqref{eq:fourier_modes} of the symplectic current itself and by integrating the condition \eqref{eq:div_free} we can express,
\begin{align}\label{eq:sympl_curre_corr}
\left.P^r_{\delta_1,\delta_2}\right|_{r=\infty}=\left.P^r_{\delta_1,\delta_2}\right|_{r=0}+B_{\delta_1,\delta_2}\,,
\end{align}
where we have set,
\begin{align}
B_{\delta_1,\delta_2}=i\,\int_{0}^\infty dr\,\left(-w\,\left(S^\prime\,P^r_{\delta_1,\delta_2}+P^t_{\delta_1,\delta_2}\right)+k_i\,P^i_{\delta_1,\delta_2} \right)\,.
\end{align}
This shows that when e.g. one of the two perturbations $\delta_2$ is chosen to be hydrodynamic mode $\delta_H$ we can expand in $\varepsilon$ to obtain,
\begin{align}
P^\mu_{\delta_1,\delta H}&=\varepsilon\,P^{\mu (1)}_{\delta_1,\delta H}+\varepsilon^2\,P^{\mu (2)}_{\delta_1,\delta H}+\cdots\,.
\end{align}
The leading term $P^{r(1)}_{\delta_1,\delta H}$ is the symplectic current formed by the $\varepsilon$ terms in the hydrodynamic expansion \eqref{eq:hydro_pert_exp}. As such, it will not offer us additional information apart from facts about the static perturbations since $\delta_1$ always refers to static perturbations of the background. At second order in the $\varepsilon$ expansion, we have,
\begin{align}\label{eq:sympl_curre_diss_corr}
\left.P^{r(2)}_{\delta_1,\delta_2}\right|_{r=\infty}=\left.P^{r(2)}_{\delta_1,\delta_2}\right|_{r=0}+B^{r(2)}_{\delta_1,\delta_2}\,,
\end{align}
with,
\begin{align}
B^{(2)}_{\delta_1,\delta_2}=i\,\int_{0}^\infty dr\,\left(-\omega\,\left(S^\prime\,P^{r(1)}_{\delta_1,\delta_2}+P^{t(1)}_{\delta_1,\delta_2}\right)+q_i\,P^{i(1)}_{\delta_1,\delta_2} \right)\,.
\end{align}
We therefore see that bulk integrals will in principle be present in our final expressions. However, after moving to a fluid frame in which the transport coefficients are of physical significance and using properties of the static perturbations, the bulk integral will disappear from the final constitutive relations.

Before we embark on our journey to fix the dissipative parts of the constitutive relations \eqref{eq:const_rels}, we would like to point at some piece of information that will be useful and that we can obtain from the symplectic current formed entirely from our static solutions. By considering the symplectic current $P^\mu_{\delta v^i,\delta m^j}$ and applying equation \eqref{eq:sympl_curre_corr} we obtain,
\begin{align}
\delta f_g^{(1)}=-\frac{4\pi \varrho_h}{s}\,\delta f_b^{(0)}+\frac{4\pi \varrho_n}{s}\,.
\end{align}
In Appendix \ref{app:static_constraints} we list all the constraints we can obtain by forming the symplectic current of all possible combinations of the static perturbations in subsections \ref{sec:pert_thermo} and \ref{sec:pert_diffeo}. In contrast to the case of normal fluids that were studied in \cite{Donos:2022uea}, these bulk constraints will be used in order to show all the transport coefficients are fixed by the event horizon of our thermal states \eqref{eq:background}.

The static perturbations of subsection \ref{sec:pert_diffeo} which contain sources for the asymptotic metric, will let us express the boundary stress tensor in terms of horizon data and bulk integrals of the perturbation. Similarly, the perturbation $\delta m_a$ discussed in subsection \ref{sec:pert_thermo} will let us do the same for the electric current. As we will see, this will leave us with constants of integration on the horizon which still need to be expressed in terms of our hydrodyanmic variables. Similarly to \cite{Donos:2022uea}, we will achieve this by considering the symplectic current formed by using the hydrodynamic mode \eqref{eq:hydro_pert_exp} along with the temperature variations $\delta_T$ and boosts $\delta v_i$ of subsection \ref{sec:pert_thermo}.

In order to read off the corrections $\delta\langle T^{ab}\rangle_{(2)}$, we will consider the symplectic current $P^\mu_{\delta s_{ab},\delta_H}$ to obtain,
\begin{align}\label{eq:stress_tensor_horizon}
\varepsilon\,\delta\langle T^{tt} \rangle_{(2)}=&\frac{i}{4 \pi}\left(-q_i s\,\left(\delta s^{it}+\delta v^i\right)+\omega\left(s-\mu_t\, \xi\right)\,\delta_{ij}\delta s^{ij}+\omega\frac{c_\mu}{s T}\left(s-\mu_t \,\xi\right)\delta T\right.\nn
&+\omega\,\frac{\xi}{s}\,\left(s-\mu_t\, \xi\right)\,\delta\mu+\omega\mu_t\,\xi \delta s^{tt}+\varepsilon\,4\pi i\, \partial_\mu \varrho_h\, \mu_t\, \delta b_t^{(2)(0)}-\varepsilon\,8i\,\pi^2\, T\,\delta^{ij}\delta g_{ij}^{(2)(0)}\nn
&+\varepsilon\,2\pi i\, T\,\mu_t\,\xi\,\left(\delta g_{rr}^{(2)(0)}+\delta g_{tt}^{(2)(0)}\right)+s\mu_t\,\omega \left(\partial_T \rho^{(0)}\partial_\mu \rho^{(0)}+\partial_T \phi^{(0)}\partial_\mu \phi^{(0)}\right)\,\delta T\nn
&\left.+s \mu_t\, \omega\,\left(\left(\partial_\mu \rho^{(0)}\right)^2+\left(\partial_\mu \phi^{(0)}\right)^2\right)\,\delta \mu\right)+\varepsilon\,B^{(2)}_{\delta_{s_{tt}},\delta_H}\,,\nn
\varepsilon\,\delta\langle T^{ti} \rangle_{(2)}=&-\varepsilon\,\frac{4\pi}{s}\left(\mu_t \varrho_n+T s\right)\delta g_{ti}^{(2)(0)}+ i \omega \mu_t (\delta f_b^{(0)})^2\left(\delta m^i+\mu_t \delta v^i\right)+\varepsilon\,B^{(2)}_{\delta_{s_{ti}},\delta_H}\,,\nn
\varepsilon\,\delta\langle T^{ij} \rangle_{(2)}=&-\frac{i s}{4\pi}\left(2\left(q^{(i}\delta s^{j) t}+q^{(i}\delta v^{j)}\right)-\delta^{ij} q_k\left(\delta s^{kt}+\delta v^k\right)\right)\nn
&-\frac{i \omega}{4 \pi}\left(\delta^{ij}\left(2\, s\, \delta_{kl}\delta s^{kl}-s \,\delta s^{tt}+\frac{c_\mu}{T}\delta T+\xi\, \delta\mu\right)-2 s\, \delta s^{(ij)}\right)\nn
&-\varepsilon\,\delta^{ij}\frac{s\, T}{2}\left(\delta g_{rr}^{(2)(0)}+\delta g_{tt}^{(2)(0)}\right)-\varepsilon\,\varrho_h\delta^{ij}\, \delta b_t^{(2)(0)}+\varepsilon\,B^{(2)}_{\delta_{s_{ij}},\delta_H}\,.
\end{align}
By considering the symplectic current $P^\mu_{\delta s_{it},\delta_H}$, we can obtain the form,
\begin{align}\label{eq:tti_alt_form}
\varepsilon\,\delta\langle T^{ti} \rangle_{(2)}=&-\frac{i}{4\pi T}\left(q^i\,\left(s T\delta^{kl}\delta s_{kl}+c_\mu \delta T+T\xi\delta\mu\right)+\omega\,\delta v^i\,\left(s T+\mu \varrho_n\right)\right)\nn
&-\frac{i}{4\pi T}\left(\omega\,\varrho_n\,\delta m^i-\delta f_b^{(0)} \varrho_h\left(\delta m^i+\mu \delta v^i\right)\omega \right)+\varepsilon\,\frac{s}{4 \pi}\delta^{ij}\left(\delta g_{tj}^{(2)(1)}-2\delta g_{tj}^{(2)(0)}g_{(1)}\right)
\nn
&+\varepsilon\,\frac{is}{4\pi }\omega \left(\delta m^i+\mu \delta v^i\right)\delta f_c^{(1)}+\varepsilon\,\varrho_h\delta^{ij} \delta b^{(2)(0)}_j+\varepsilon\,B^{(2)}_{\delta_{s_{it}},\delta_H}\,,
\end{align}
This form for $\delta\langle T^{ti} \rangle_{(2)}$ is equivalent to the one we listed in \eqref{eq:stress_tensor_horizon} because $P^\mu_{\delta s_{it},\delta_H}$ and $P^\mu_{\delta s_{ti},\delta_H}$ linearly combine to $P^\mu_{\delta v_i,\delta_H}$ which is source free.

The above expressions express the first dissipative correction to the ideal superfluid as functions of the our hydrodynamic variables as well as the horizon constants of integration $\delta^{ij}\delta g_{ij}^{(2)(0)}$ and $\delta b_t^{(2)(0)}$ for the correction $\delta f^{(2)}$ in our hydrodynamic expansion \eqref{eq:hydro_pert_exp}. Moreover, we have the appearance of integrals over the bulk which, as we will later see, are artifacts of the fluid frame. Moreover, the appearance of the constants of integration $\delta g_{tt}^{(2)(0)}+\delta g_{rr}^{(2)(0)}$ and $\delta g_{ti}^{(2)(0)}$ should not bother us as they can be considered as corrections to the local temperature $\delta T$ and fluid velocity $\delta v_i$ respectively. In other words, we will be above to absorb them in the definition of $\delta T$ and $\delta v_i$ via a change of fluid frame.

In order to obtain similar expressions for the leading dissipative corrections $\delta\langle J^a\rangle$ of the electric current, we will apply equation \eqref{eq:sympl_curre_diss_corr} for the symplectic current $P^\mu_{\delta_{m_a},\delta_H}$. This allows us to write,
\begin{align}\label{eq:el_current_horizon}
\varepsilon\,\delta\langle J^t\rangle_{(2)}=&-\frac{i\omega}{4\pi}\xi\left(\delta^{kl}\delta s_{kl}-\delta s^{tt}\right)-\frac{i\omega}{4\pi s}\xi\left(\frac{c_\mu}{T}\delta T+\xi\delta \mu\right)\nn
&+\frac{i\omega s}{4\pi}\left(\left( \partial_T \rho^{(0)}\partial_\mu\rho^{(0)}+\partial_T \phi^{(0)}\partial_\mu\phi^{(0)}\right)\delta T+ \left(\left(\partial_\mu\rho^{(0)}\right)^2+\left(\partial_\mu\phi^{(0)}\right)^2\right)\delta\mu\right)\nn
&-\varepsilon\,\partial_\mu \varrho_h \,\delta b^{(2)(0)}_t-\varepsilon\,\frac{T \xi}{2}\left(\delta g_{rr}^{(2)(0)}+\delta g_{tt}^{(2)(0)}\right)+\varepsilon\,B^{(2)}_{\delta_{m_{t}},\delta_H}\,,\nn
\varepsilon\,\delta\langle J^i\rangle_{(2)}=&i\omega \left( \delta f_b^{(0)}\right)^2\,\left(\delta m^i+\mu_t \delta v^i\right)-\varepsilon\,\frac{4\pi}{s}\varrho_n \delta^{ij}\delta g_{tj}^{(0)}+\varepsilon\,B^{(2)}_{\delta_{m_{i}},\delta_H}\,.
\end{align}
Once again, the above expressions contain constants of integration for the perturbation which need to be fixed in terms of our hydrodynamic variables.

To achieve this, we will consider equation \eqref{eq:sympl_curre_diss_corr} for the symplectic currents $P^\mu_{\delta v^i,\delta_H}$ and $P^\mu_{\delta T,\delta_H}$ which don't contain any sources. Moreover, we will examine the radial component of the vector field equation of motion in \eqref{eq:eom} close to the black hole horizon. This step was also necessary in \cite{Donos:2021pkk} and in \cite{Donos:2022uea} in order to eliminate the horizon degree of freedom.

By considering the symplectic current $P^\mu_{\delta v^i,\delta H}$, we obtain the equation,
\begin{align}
&i\omega\left(\delta f_b^{(0)}\right)^2\mu_t \left(\delta m_i+\mu_t \delta v_i\right)-\varepsilon\,\varrho_h \delta b^{(2)(0)}_i-\varepsilon\,\frac{4\pi}{s}\left(s\, T+\mu_t \varrho_n\right)\delta g_{ti}^{(2)(0)}
\nn
&+\frac{i}{4\pi T}\left( q_i\left(s \,T\,\delta^{kl}\delta s_{kl}+c_\mu \delta T+T \xi\, \delta\mu\right)+\omega\left(s T+\mu_t \varrho_n\right)\delta v_i\right)\nn
&+\frac{i}{4\pi T}\left(\omega \varrho_n\,\delta m_i-\delta f_b^{(0)} \varrho_h \,\omega\left(\delta m_i+\mu_t \delta v_i\right)\right)\nn
&-\varepsilon\,\frac{s}{4\pi}\left(\delta g_{ti}^{(2)(1)}-2\delta g_{ti}^{(2)(0)}g_{(1)}\right)-\frac{is}{16\pi^2 T}\omega\,\delta f_c^{(1)}\left(\delta m_i+\mu_t \delta v_i\right)=\varepsilon\,B^{(2)}_{\delta_{v^i},\delta H}\,.
\end{align}
The above can used to show the equivalence between the expressions \eqref{eq:tti_alt_form} and the one given in \eqref{eq:stress_tensor_horizon} after noting that,
\begin{align}
B^{(2)}_{\delta_{v_i},\delta H}=B^{(2)}_{\delta_{s_{ti}},\delta H}-B^{(2)}_{\delta_{s_{it}},\delta H}\,.
\end{align}
Our next step is to consider equation \eqref{eq:sympl_curre_diss_corr} for the symplectic current $P^\mu_{\delta T,\delta H}$ yielding,
\begin{align}\label{eq:constr_deltaT_deltaH}
&\frac{c_\mu^2}{s T^2}\,\omega\, \delta T+\frac{c_\mu \xi}{s T}\,\omega\, \delta\mu+\frac{c_\mu}{T}\omega\left(\delta^{ij}\delta s_{ij}-\delta s^{tt}\right)-\varepsilon\,2\pi i\,c_\mu\left(\delta g_{rr}^{(2)(0)}+\delta g_{tt}^{(2)(0)}\right)\nn
&-\varepsilon\,4\pi i\,\partial_T \varrho_h \, \delta b^{(2)(0)}_t-\varepsilon\,8\,i \,\pi^2\,\delta^{ij} \delta g_{ij}^{(2)(0)}-s\omega \left(\left(\partial_T\rho^{(0)}\right)^2+\left(\partial_T\phi^{(0)}\right)^2\right)\delta T\nn
&-s\,\omega\left(\partial_T \rho^{(0)}\partial_\mu \rho^{(0)}+\partial_T \phi^{(0)}\partial_\mu \phi^{(0)}\right) \delta\mu=\varepsilon\,\pi i\, B^{(2)}_{\delta_T,\delta H}\,.
\end{align}
The above equation in combination with the radial component of the vector equation in \eqref{eq:eom} evaluated on the horizon,
\begin{align}\label{eq:vector_eom_hor}
\varepsilon\,\delta b^{(2)(0)}_t=\frac{4\pi i}{s\, q_e^2 \left(\rho^{(0)}\right)^2}\left(\varrho_h q_i \left(\delta s^{it}+\delta v^i\right)-\omega \left(\varrho_h \delta^{ij}\delta s_{ij}+\partial_T \varrho_h \delta T+\partial_\mu \varrho_h \delta\mu\right) \right)\,,
\end{align}
can be used to eliminate the constants $\delta^{ij}\delta g_{ij}^{(2)(0)}$ and $\delta b_t^{(2)(0)}$ from the constitutive relations \eqref{eq:stress_tensor_horizon} and \eqref{eq:el_current_horizon}. By eliminating those we can express all our constitutive relations in terms of the hydrodynamic variables $\delta T$, $\delta v_i$ and $\delta\theta_{(v)}$. However, there is still a set of bulk integrals which are not equal to zero. By performing the change of frame given by,
\begin{align}
\delta T&\to \delta T-\frac{\varepsilon}{c_\mu+\mu_t\,\xi}\left( B^{(2)}_{\delta_{s_{tt}},\delta H}-T\,B^{(2)}_{\delta_T,\delta H}\right)\,,\nn
\delta v^i&\to \delta v^i-\frac{\varepsilon}{s\,T+\mu_t\,\varrho_n}B^{(2)}_{\delta_{s_{ti}},\delta H} \,,
\end{align}
we can eliminate the bulk integrals from the dissipative corrections $\delta\langle T^{ti} \rangle_{(2)}$ and $\delta\langle T^{tt} \rangle_{(2)}$. However, the change of fluid frame will make them appear in the dissipative corrections $\delta\langle T^{ij} \rangle_{(2)}$, $\delta\langle J^t\rangle_{(2)}$ and $\delta\langle J^i\rangle_{(2)}$. The important observation is that the combinations,
\begin{align}\label{eq:bulk_int_relations}
-\frac{s}{c_\mu+\mu_t\,\xi}\left( B^{(2)}_{\delta_{s_{tt}},\delta H}-T\,B^{(2)}_{\delta_T,\delta H}\right)\delta^{ij}+B^{(2)}_{\delta_{s_{ij}},\delta H}&=0\,,\nn
-\frac{\xi}{c_\mu+\mu_t\,\xi}\left( B^{(2)}_{\delta_{s_{tt}},\delta H}-T\,B^{(2)}_{\delta_T,\delta H}\right)+B^{(2)}_{\delta_{m_{t}},\delta H}&=0\,,\nn
-\frac{\varrho_n}{s\,T+\mu_t\,\varrho_n}B^{(2)}_{\delta_{s_{ti}},\delta H}+B^{(2)}_{\delta_{m_{i}},\delta H}&=0\,,
\end{align}
that would appear, are actually equal to zero. This is a non-trivial result which in order to be shown, requires the use of the constraints that we list in Appendix \ref{app:static_constraints} along with equations of motion of the ideal fluid \eqref{eq:ideal_eoms}.

After cancelling out the bulk integrals, we land in a non-standard fluid frame which we need to change in order to bring our constitutive relations to a more conventional form. In the next subsection we will rewrite our theory of hydrodynamics in the so called transverse frame. This will help us read off the transport coefficients without resorting to the relevant Kubo formulae.

\subsection{The Transverse Fluid Frame}\label{sec:fluid_frames}

In this subsection we wish to bring our constitutive relations to a form compatible with a well established frame in the literature. In the transverse frame that we wish to consider, the constitutive relations of our system can be decomposed to ideal and dissipative pieces according,
\begin{align}\label{eq:const_rels_trans}
\langle T^{ab}\rangle&=\langle T^{ab}\rangle_{ideal}+\langle T^{ab}\rangle_{diss}\,,\nn
\langle J^{a}\rangle&=\langle J^{a}\rangle_{ideal}+\langle J^{a}\rangle_{diss}\,,
\end{align}
with the perturbations of the ideal parts being given by the $\mathcal{O}(\varepsilon)$ terms of equations \eqref{eq:const_rels}. The constraints that fix the transverse frame that we wish to consider are,
\begin{align}\label{eq:trans_constr}
u_a\,\langle T^{ab}\rangle_{diss}=0\,,\quad
u_a\,\langle J^{a}\rangle_{diss}=0\,.
\end{align}
In a spacetime of $d+1$ spacetime dimensions, the above equations constitute $d+2$ constraints. These can be achieved by transforming the local temperature $\delta T$, chemical potential $\delta\mu$ and fluid velocity $\delta v^a$. Given the fact that the normal fluid velocity, $u_a$ satisfies the normalisation condition $u^2=-1$, we have exactly $d+2$ variables to achieve the conditions \eqref{eq:trans_constr}.

It can be shown that the most general form for the constitutive relations after imposing these constraints takes the form,
\begin{align}\label{eq:const_rels_transverse}
\langle T_{ab}\rangle_{diss}=&-\eta\,\sigma^n_{ab}-\zeta_1\,P_{ab}\,\nabla_c u^c-\zeta_2\,P_{ab}\,\nabla_c\left(\varrho_s n^c\right)\,,\nn
\langle J_{a}\rangle_{diss}=&-\sigma\,P_{ab}\,\left(\nabla^b\left(\frac{\mu}{T}\right)-\frac{1}{T}\,F^{bc}\,u_c\right)\,,
\end{align}
where we have defined,
\begin{align}
\sigma^n_{ab}=&P_a{}^{c}P_{b}{}^{d}\left(\nabla_c u_d+\nabla_d u_c \right)-P_{ab}\,P^{cd}\,\nabla_c u_d\,,\nn
n_a=&-\frac{1}{\mu_s}\,P_a{}^b\,m_b\,,\nn
\mu_s=&u^a\,m_a\,,
\end{align}
and the projection operator is as usual $P_{ab}=g_{ab}+u_a\,u_b$. In order to transform to this frame, we need to perform a redefinition of the local chemical potential to obtain a Josephson relation,
\begin{align}\label{eq:mu_diss}
\mu=u^a\,m_a+\mu_{diss}\,,
\end{align}
with dissipative corrections given by,
\begin{align}
\mu_{diss}=\zeta_3\,\nabla_c\left( \varrho_s\,n^c\right)+\zeta_2\,\nabla_c u^c\,.
\end{align}
Notice that even though the components $\langle T_{tt}\rangle_{diss}$ and $\langle J_{t}\rangle_{diss}$ are trivial, the fact that the chemical potential \eqref{eq:mu_diss} contains dissipative terms it is enough for the ideal parts to lead to dissipation.

By linearising the expression we have in fluid and source perturbations, we obtain the constitutive relations,
\begin{align}\label{eq:const_rels_pert_transverse}
\delta \langle T^{ij}\rangle_{diss}=&\eta\left( -2\left(\partial^{(i}\delta s^{j)t}+\partial^{(i}\delta v^{j)}\right)+\delta^{ij} \partial_k \left(\delta s^{kt}+\delta v^k\right)-2\,\partial_t\, \delta s^{(ij)}+\delta^{ij}\,\delta^{kl}\partial_t\delta s_{kl}\right)\nn
&-\zeta_1 \delta^{ij}\left(\partial_k\left(\delta s^{kt}+\delta v^k\right)+\partial_t\delta s_{kl}\delta^{kl}\right)+\delta^{ij} \zeta_2\, \chi_{JJ}\,\partial_k\,\left(\delta m^k+\mu_t \, \delta v^k\right)\,,\nn
\delta \langle J^i\rangle_{diss}=&\frac{\sigma}{T} \left( \partial^i \left(\delta m_t-\delta \mu\right)+\frac{\mu_t}{T}\,\partial^i \delta T-\partial_t\delta m^i\right)\,,
\end{align}
with the linear piece of the dissipative part of the chemical potential reading,
\begin{align}\label{eq:mu_diss_pert}
\delta\mu_{diss}=\zeta_2\,\left( \partial_i \left(\delta s^{it}+\delta v^i\right)+\partial_t\delta^{ij}\delta s_{ij}\right)-\,\zeta_3\,\chi_{JJ}\,\partial_i\left(\delta m^i+\mu_t\, \delta v^i\right)\,.
\end{align}

In order to compare to the above form for the constitutive relations, we can use the relations \eqref{eq:constr_deltaT_deltaH} and \eqref{eq:vector_eom_hor} to express the constitutive relations \eqref{eq:stress_tensor_horizon} and \eqref{eq:el_current_horizon} entirely in terms of our hydrodynamic variables. The relations \eqref{eq:bulk_int_relations} guarantee that the bulk integrals are a frame artifact and that they can be removed from our constitutive relations. After transforming to the transverse frame, our constitutive relations take the form \eqref{eq:const_rels_trans} with dissipative parts \eqref{eq:const_rels_pert_transverse} and \eqref{eq:mu_diss_pert} fixed by the incoherent conductivity and shear viscosity,
\begin{align}\label{eq:sigma_eta}
\sigma&=\frac{s^2 T^3}{(s T+\mu_t \varrho_n)^2}\,(\delta f_b^{(0)})^2 \,\tau^{(0)}\,,\nn
\eta&=\frac{s}{4\pi}\,.
\end{align}
In order to express the three bulk viscosities $\zeta_i$ in terms of horizon data and thermodynamic quantities, it is convenient to make the change of background thermodynamic variables $(T,\mu_t)\rightarrow(s,\varrho)$. For  any quantity $F$, using the chain rule, the thermodynamic derivatives are connected through,
\begin{align}
\partial_{\mu_t} F&=\chi_{QQ}\,\partial_\varrho F+\xi\,\partial_s F\,,\\
\partial_T F&=\xi\,\partial_\varrho F+\frac{c_\mu}{T}\,\partial_s F\,.
\end{align}
After performing this change of variables, the three bulk viscosities take the form,
\begin{align}\label{eq:bulk_viscosities}
\zeta_1=&\frac{s}{4\pi}\left( \left(s\,\partial_s \phi^{(0)}+\varrho\, \partial_\varrho \phi^{(0)}\right)^2+\left(s\,\partial_s \rho^{(0)}+\varrho\, \partial_\varrho \rho^{(0)}\right)^2\right)\nn
&\qquad+\frac{4\pi}{s\,q_e^2\left(\rho^{(0)}\right)^2}\left(\varrho_h-\varrho\, \partial_\varrho \varrho_h-s\,\partial_s \varrho_h\right)^2\,,\nn
\zeta_2=&\frac{s}{4\pi}\left(\partial_\varrho \phi^{(0)}\left(s\,\partial_s \phi^{(0)}+\varrho\, \partial_\varrho \phi^{(0)}\right)+\partial_\varrho \rho^{(0)}\left(s\,\partial_s \rho^{(0)}+\varrho \,\partial_\varrho \rho^{(0)}\right)\right)\nn
&\qquad-\frac{4\pi}{s\,q_e^2\left(\rho^{(0)}\right)^2}\partial_\varrho \varrho_h \left(\varrho_h-\varrho\, \partial_\varrho \varrho_h-s\,\partial_s \varrho_h\right)\,,\nn
\zeta_3=&\frac{s}{4\pi}\left(\left( \partial_\varrho \phi^{(0)}\right)^2+\left( \partial_\varrho \rho^{(0)}\right)^2\right)+\frac{4\pi}{s\,q_e^2\left(\rho^{(0)}\right)^2} \left(\partial_\varrho \varrho_h\right)^2\,,
\end{align}
where $\varrho_h$ is the horizon charge density we defined in equation \eqref{eq:hcharge_density}. From these expressions  obviously $\zeta_1>0$. Also, using the Schwarz inequality we can show that\footnote{This corrects a typo in the relation previously given in \cite{Herzog:2011ec} which was based on the positivity of entropy productioon.} $\zeta_1\zeta_3\geq \zeta_2^2$.

\section{Limits}\label{sec:limits}

In this following subsections we will consider the limit of our hydrodynamics close to the phase transition as well as at zero chemical potential. This is possible because of our explicit expressions for the five transport coefficients in terms of horizon data.

\subsection{Near Critical Point}\label{sec:limits_tc}
Here, we wish to consider the limit of the hydrodynamic modes close to the phase transition as we approach it from the broken phase. This is certainly possible since we have the holographic expressions \eqref{eq:sigma_eta} and \eqref{eq:bulk_viscosities} for the transport coefficients in terms of thermodynamics and horizon data.

In order to take the limit close to the critical point, we can vary either the background chemical potential $\mu_t$, the temperature $T$ or the scalar deformation parameter $\phi_{(s)}$. For simplicity, we will only consider variations of the temperature away from its critical value $T_c$ for fixed chemical potential and scalar deformation. Setting $\delta\lambda=T_c-T$, we can write the expansion,
\begin{align}
\rho=\delta\lambda\,\rho_{0}+\delta\lambda^3\,\rho_{1}+\cdots,
\end{align}
for the bulk scalar dual to the amplitude of the complex order parameter. A simple analysis of the gauge field equation of motion in\eqref{eq:eom} reveals that close to the transition we can write,
\begin{align}
\chi_{JJ}=\frac{c_{JJ}}{\delta\lambda^2}+\cdots\,,\quad \varrho_h=\varrho+\delta\lambda^2\,\varrho_{h 1}+\cdots\,,\qquad \varrho_n=\varrho+\delta\lambda^2\,\varrho_{1}+\cdots\,.
\end{align}\label{eq:}
The above relations yield the singular limit for the bulk viscosities,
\begin{align}\label{eq:bv_blowup}
\zeta_i=\frac{c_i}{\delta\lambda^2}+\cdots\,,
\end{align}
with $c_i$ remaining finite at the transition.

The first task is to find the hydrodynamic modes of our superfluid away from the critical point and then take the limit. More concretely, we would like to turn of the sources $\delta s_{ab}$ for the metric and the gauge field $\delta\mu_a$ and solve the Ward identities \eqref{eq:Ward} given the constitutive relations in the transverse frame that we discussed in section \ref{sec:fluid_frames}. Without loss of generality, we can take the wavevector of the fluctuation to lie entirely on the $x^1$ axis with components $k_1=\varepsilon\,q$ and $k_2=0$ with $\varepsilon$ the hydrodynamic expansion parameter. The goal is to fix the dispersion relations of the relevant quasi-normal modes whose frequency we can expand according to,
\begin{align}
\omega=\omega_{[1]}\,\varepsilon+\omega_{[2]}\,\varepsilon^2+\cdots\,.
\end{align}

In order to find the modes we are after, we Fourier expand our hydrodynamic variables according to,
\begin{align}\label{eq:hydro_exp}
\delta T&=e^{-i\omega\,t+i\varepsilon q x^1}\,\varepsilon\left( \delta T_{0}+\varepsilon\,\delta T_{1}+\cdots\right)\,,\nn
\delta \theta_{(v)}&=e^{-i\omega\,t+i\varepsilon q x^1}\,\left( \delta \theta_{0}+\varepsilon\,\delta \theta_{1}+\cdots\right)\,,\nn
\delta v^i&=e^{-i\omega\,t+i\varepsilon q x^1}\,\varepsilon\left( \delta v^i_{0}+\varepsilon\,\delta v^i_{1}+\cdots\right)\,.
\end{align}
The leading part of the modes along with $\omega_{[1]}$ are determined by the ideal superfluid equations of motion \eqref{eq:ideal_eoms}. This leads to a linear algebraic system for the constants $\delta T_0$, $\delta\theta_0$ and $\delta v_0^i$ which is trivial unless the frequency satisfies the zero determinant condition,
\begin{align}
&\omega_{[1]}\left(\left(s\,T+\mu_t\,\varrho_n \right)\left(T\,\xi^2-c_\mu\,\chi_{QQ} \right)\omega_{[1]}^4-s^2\,T\,\chi_{JJ}\,q^4\right.\nn
&\qquad \left.+\left(c_\mu\,\varrho\,\varrho_n+s\,T\left(s\,\chi_{QQ} +\left(c_\mu+\mu_t\,\xi \right)\chi_{JJ}-\left(\varrho+\varrho_n \right)\xi\right) \right)\omega_{[1]}^2\,q^2\right)=0\,.
\end{align}
By solving the above equation for $\omega_{[1]}$, we fix the leading piece of the dispersion relations we are after. The first solution trivially yields,
\begin{align}
\omega^{shear}_{[1]}=0,
\end{align}
which is a transverse mode with $\delta v_0^1=0$. This is essentially the shear mode describing the diffusion of momentum along a transverse direction.

The remaining four modes come in two pairs corresponding to the first and the second sounds modes of the superfluid. Here, we will only be interested in the limit of the dispersion relations near the critical point. For the first sound we have the asymptotic behaviour,
\begin{align}
(\omega^{f.s.}_{[1]})^2=\frac{Z}{\left(s\,T+\mu_t\,\varrho\right)\left(c_\mu\,\chi_{QQ}-T\,\xi^2\right)}\,q^2+\cdots\,,
\end{align}
where for convenience we have defined the quantity,
\begin{align}
Z=s^2\,T\,\chi_{QQ}+c_\mu\,\varrho^2-2 s\,T\,\xi\,\varrho\,.
\end{align}
leading to a finite speed of sound in the small $\delta\lambda$ limit. In contrast, for the second sound we have the leading piece of the dispersion relation,
\begin{align}\label{eq:ss_speed}
(\omega^{s.s.}_{[1]})^2=\frac{\chi_{JJ}\,s^2\,T}{Z}\,q^2+\cdots
\end{align}
with a speed of sound approaching zero close to the critical point.

After imposing the vanishing of the determinant of the leading order linear system of equations, the allowed modes are determined by the kernel of the linear operator multiplying the vector with components $\delta_0$, $\delta\theta_0$ and $\delta v_0^i$. By expanding the Ward identities up to order $\varepsilon^3$, we can determine the higher order constants in the hydrodynamic expansion \eqref{eq:hydro_exp}. However, in order for this to be possible we find that the $\omega_{[2]}$ part of the frequency expansion needs to satisfy an algebraic condition which completely fixes it in terms of $\omega_{[1]}$.

More specifically, for the shear mode we have the correction,
\begin{align}
\omega^{shear}_{[2]}=-i\frac{\eta}{T\,s+\mu_t\,\varrho}q^2\,,
\end{align}
leading to a purely diffusive mode, as expected. For the leading correction to the first sound mode we obtain,
\begin{align}
\omega^{f.s.}_{[2]}=-\frac{i}{2}\frac{\zeta_1}{s\,T+\mu_t\,\varrho}q^2+\cdots\,,
\end{align}
showing the the attenuation blows up close to the critical point for fixed wavenumber. This is due the fact that the bulk viscosity of relativistic fluids blows up close to the critical point when approached from the broken phase. This has been discussed before in the context of simpler systems \cite{Donos:2022uea} with scalar order parameters.

However, for the second sound we find the limiting behaviour,
\begin{align}\label{eq:ss_attenuation}
\omega^{s.s.}_{[2]}=&-\frac{i}{2}\left(\zeta_3\,\chi_{JJ}+\frac{\zeta_1\,\chi_{JJ}\left(s\,T\,\xi-c_\mu\,\varrho\right)^2}{Z^2}\right.\nn
&\qquad\qquad\left.+\frac{2\,T^2\,\zeta_2\,\chi_{JJ}\left(s\,T\,\xi-c_\mu\,\varrho\right)+\left( s\,T+\mu_t\,\varrho\right)^2\,\sigma}{Z\,T^2} \right)q^2+\cdots
\end{align}
for the leading attenuation part showing that it remains finite close to the phase transition eventhough the bulk viscosities seem to blow up close to the transition according to \eqref{eq:bv_blowup}. This is in parallel to the observation first made in \cite{Donos:2021pkk} for holographic superfluids at zero chemical potential. In fact, by taking the zero chemical potential limit, we can match the limiting behavious of the attenuation \eqref{eq:ss_attenuation} to the expression of \cite{Donos:2021pkk}.

\subsection{Zero Chemical Potential}\label{sec:limits_mu_zero}

It is interesting to consider the zero chemical potential limit of our hydrodynamics and compare with our results in \cite{Donos:2021pkk}. For the classes of holographic theories we consider, it is easy to see that the background scalar fields $\phi$ and $\rho$  are even under changing the sign of either the chemical potential or the charge density $\varrho$. This suggests that at zero chamical potential and therefore zero charge density, we must have vanishing $\partial_\varrho \phi^{(0)}$ and $\partial_\varrho \rho^{(0)}$. Moreover, the horizon charge density $\varrho_h$ and the thermodynamic susceptibity $\xi$ are identically zero for any value of the entropy of the system. This shows that $\zeta_2=0$ while $\zeta_1$ becomes the bulk viscosity of relativistic holographic fluids \cite{Eling:2011ms, Donos:2022uea}. The final expressions for the bulk viscosities in the zero chemical potential limit are,
\begin{align}
\zeta=\zeta_1&=\frac{s^3}{4\pi}\left( \left( \partial_s\phi^{(0)}\right)^2+\left( \partial_s\rho^{(0)}\right)^2\right)=\frac{s}{4\pi}\left(\frac{T s}{c_\mu}\right)^2\left( \left( \partial_T\phi^{(0)}\right)^2+\left( \partial_T\rho^{(0)}\right)^2\right)\,,\nn
\zeta_2&=0\,,\nn
\zeta_3&=\frac{2\pi}{s\,q_e^2\,\left(\rho^{(0)}\right)^2} \left(\partial_\varrho \varrho_h\right)^2=\frac{2\pi}{s\,q_e^2\left(\rho^{(0)}\right)^2\chi_{QQ}^2} \left(\partial_\mu \varrho_h\right)^2\,.
\end{align}

In order to compare with the constitutive relations of \cite{Donos:2021pkk} for the electric current, we will express our constitutive relations in terms of the phase $\delta\theta_{(v)}$. Combining our expressions \eqref{eq:const_rels_trans} along with \eqref{eq:const_rels_transverse}, \eqref{eq:const_rels} and \eqref{eq:mu_diss_pert} we obtain,
\begin{align}
\delta\langle J^t\rangle&=\chi_{QQ}\,\delta\mu=\chi_{QQ}\,\partial_t\delta\theta_{(v)}-\chi_{QQ}\,\chi_{JJ}\,\zeta_{3}\,\partial_i\partial^i\,\delta\theta_{(v)}=\chi_{QQ}\,\partial_t\delta\theta_{(v)}-\chi_{QQ}^2\,\zeta_{3}\,\partial_t^2\,\delta\theta_{(v)}\,,\nn
\delta\langle J^i\rangle&=-\chi_{JJ}\,\partial^i\theta_{(v)}-\frac{\sigma}{T}\,\partial^i\partial_t\theta_{(v)}\,
\end{align}
where in the first line we used the ideal superfluid equations of motion \eqref{eq:ideal_eoms}. We see that the above agrees with \cite{Donos:2021pkk} after matching $\Xi=\chi_{QQ}^2\,\zeta_3$ and $\sigma_d=\sigma/T$, noting that $\delta c=q_e\,\delta\theta_{(v)}$. To make the comparison more precise we note that for small chemical potential $\left.\delta\varrho_h\right|_{here}=\left.e^{2 g^{(0)}} \tau^{(0)} a_t^{(0)}\,\delta\mu_t\right|_{there}$. For the thermodynamic supercurrent perturbation we have $\left.\delta f_b^{(0)}\right|_{here}=\left. a_x^{(0)}\right|_{there}$. Moreover, due to different normalisation of the bulk scalar we have $\left.\rho^{(0)}\right|_{here}=\sqrt{2}\,\left.\rho^{(0)}\right|_{there}$.

\section{Discussion}\label{sec:discussion}

In this paper we have used the techniques that were recently developed in \cite{Donos:2021pkk,Donos:2022xfd,Donos:2022uea} to study the hydrodynamic limit of fluctuations in a holographic superfluid at finite chemical potential. Based on general arguments, we expected organise the long wavelength limit of the stress tensor and the electric current in terms of a derivative expansion of appropriate hydrodynamic variables which have a clear interpretation in the infinite wavelength, thermodynamic limit. After fixing a specific fluid frame, we would then expect the leading dissipative corrections to be parametrised by a set of independent transport coefficients. For a relativistic fluid which preserves homogeneity and isotropy, we expect five independent coefficients, the incoherent conductivity $\sigma$, the shear viscosity $\eta$  and the three bulk viscosities $\zeta_i$.

An important by-product of our derivation was the explicit expressions for the dissipative transport coefficients in equations \eqref{eq:sigma_eta} and \eqref{eq:bulk_viscosities} in terms of thermodynamics and the black hole horizon data. Our results confirm \cite{Son:2007vk,Iqbal:2008by,Herzog:2011ec} that in the leading gravitational limit and while preserving translations and isotropy, the shear viscosity $\eta$ is fixed by the entropy density of the theory according to \eqref{eq:sigma_eta}. The results for the incoherent conductivity and the bulk viscosities are new as far as we know.

As an application of our results, in section \ref{sec:limits_tc} we studied the limit of the hydrodynamic fluctuations close to the critical temperature. As we saw, the bulk viscosities $\zeta_i$ blow up close to the critical, signalling the breakdown of the hydrodynamic expansion. The two longitudinal sound modes behave differently close to the critical temperature. The first sound behaves in a way similar to the behaviour that was discusses earlier in \cite{Donos:2022uea}. The speed of sound remains finite while the attenuation blows up dues to the leading behaviour of the bulk viscosity $\zeta_1$. The second sound, which is due to the superfluid component, has a vanishing speed of sound close to the transition. However, the attenuation remains finite, similarly to what happens in superfluids at zero chemical potential \cite{Donos:2021pkk}.

Given the fact that our expressions for the transport coefficients are determined by data on the black hole horizon, it is possible to consider their low temperature limit for a given ground state geometry. Large classes of holographic ground states have been considered over the years in different works \cite{Horowitz:2009ij,Gubser:2009gp,Gouteraux:2012yr} in models which are sub cases of the general model in equation \eqref{eq:bulk_action}. It would be interesting to use our results to study the effects of dissipation in holographic superfluids at low temperatures by using these geometries.

The fact that the three bulk viscosities blow up close to the phase transition is due to the amplitude mode of the superfluid becoming exactly gapless. Below the phase transition this mode becomes gapped, joining the rest of the UV modes and we can safely integrate it out. An interesting direction to pursuit further in the future is to include this mode in the hydrodynamic description and obtain an effective theory which is valid up to energy scales which are equal to the gap of this universal mode and even beyond that. This is certainly possible given the recent progress that was made in \cite{Donos:2022xfd}.

\appendix

\section{Constraints for static perturbations}\label{app:static_constraints}

In this Appendix we will list the constraints resulting from the symplectic current when constructed from the static pairs of perturbations discussed in subsections \ref{sec:pert_thermo} and \ref{sec:pert_diffeo}. In particular, given that for those pairs of perturbations only the radial component is non-trivial, the divergence free condition \eqref{eq:div_free} yields the radial constraint,
\begin{align}
P^r_{\delta_1,\delta_1}=\left.P^r_{\delta_1,\delta_1}\right|_{r=0}=\left.P^r_{\delta_1,\delta_1}\right|_{r=\infty}\,.
\end{align}
The inequivalent constraints we can obtain read,
\begin{align}
&-e^{2g}a'\delta f_b+(aa^\prime-U^\prime+2e^{2g}g^\prime)\delta f_g+aU\delta f^\prime_b+(U-e^{2g})\delta f^\prime_g=-\varrho+\mu_t\, \chi_{jj}\,,\nn
&e^{2g}a^\prime\delta f_b-2e^{2g}g^\prime\delta f_g+e^{2g}\delta f^\prime_g=\varrho-\mu_t\, \chi_{jj}\,,\nn
&-a^\prime\partial_{\mu_t}a+U\phi^\prime\partial_{\mu_t}\phi+U\rho^\prime\partial_{\mu_t}\rho+2U\partial_{\mu_t}g^\prime+\partial_{\mu_t}U^\prime=0\,,\nn
&e^{2g}\Big(2(U^\prime-2Ug^\prime-aa^\prime)\partial_{\mu_t}g-2g^\prime\partial_{\mu_t}U-U\phi^\prime\partial_{\mu_t}\phi-U\rho^\prime\partial_{\mu_t}\rho-4U\partial_{\mu_t}g^\prime-a\partial_{\mu_t}a^\prime\Big)=\xi\, T\,,\nn
&e^{2g}\Big(2(U^\prime-2Ug^\prime-aa^\prime)\partial_Tg-2g^\prime\partial_TU-U\phi^\prime\partial_T\phi-U\rho^\prime\partial_T\rho-4U\partial_Tg^\prime-a\partial_Ta^\prime\Big)=c_\mu\,,\nn
&e^{2g}\Big(-a^\prime\partial_Ta+U\phi^\prime\partial_T\phi+U\rho^\prime\partial_T\rho+2U\partial_Tg^\prime+\partial_TU^\prime\Big)=s\,,\nn
&e^{2g}\left(U^\prime-2Ug^\prime-aa^\prime\right)=s\,T\,.
\end{align}
The above equations where obtained by considering the symplectic currents in the order $P^\mu_{\delta_{v_i},\delta_{m_i}}$,  $P^\mu_{\delta_{s_{it}},\delta_{m_i}}$,  $P^\mu_{\delta_{s_{ii}},\delta_{m_t}}$,  $P^\mu_{\delta_{s_{tt}},\delta_{m_t}}$, $P^\mu_{\delta_T,2\delta_{s_{ii}}-\mu_t\,\delta_{m_t}}$, $P^\mu_{\delta_{s_{ii}},\delta T}$ and $P^\mu_{\delta_{s_{it}},\delta_{v_{i}}-\mu_t\,\delta_{m_i}}$. Finally, from the symplectic current $P^\mu_{\delta_T,\delta_{m_t}}$ we obtain the bulkier constraint,
\begin{align}
&-2e^{2g}\left(-a^\prime\partial_Ta+U\phi^\prime\partial_T\phi+U\rho^\prime\partial_T\rho+2U\partial_Tg^\prime+\partial_TU^\prime\right)\partial_{\mu_t}g\nn
&+2e^{2g}\left(-a^\prime\partial_{\mu_t}a+U\phi^\prime\partial_{\mu_t}\phi+U\rho^\prime\partial_{\mu_t}\rho+2U\partial_{\mu_t}g^\prime+\partial_{\mu_t}U^\prime\right)\partial_Tg\nn
&-e^{2g}\left(2\partial_Tg^\prime+\rho^\prime\partial_T\rho+\phi^\prime\partial_T\phi\right)\partial_{\mu_t}U+e^{2g}\left(2\partial_{\mu_t}g^\prime+\rho^\prime\partial_{\mu_t}\rho+\phi^\prime\partial_{\mu_t}\phi\right)\partial_TU\nn
&+e^{2g}U\left(\partial_T\rho^\prime\partial_{\mu_t}\rho-\partial_T\rho\partial_{\mu_t}\rho^\prime\right)+e^{2g}U\left(\partial_T\phi^\prime\partial_{\mu_t}\phi-\partial_T\phi\partial_{\mu_t}\phi^\prime\right)\nn
&+e^{2g}\left(\partial_{\mu_t}a^\prime\partial_Ta-\partial_{\mu_t}a\partial_Ta^\prime\right)=-\xi\,.
\end{align}
\newpage
\bibliographystyle{utphys}
\bibliography{refs}{}

\providecommand{\href}[2]{#2}\begingroup\raggedright\begin{thebibliography}{10}

\bibitem{Aharony:1999ti}
O.~Aharony, S.~S. Gubser, J.~M. Maldacena, H.~Ooguri, and Y.~Oz, ``{Large N
  field theories, string theory and gravity},''
  \href{http://dx.doi.org/10.1016/S0370-1573(99)00083-6}{{\em Phys. Rept.}
  {\bfseries 323} (2000) 183--386},
\href{http://arxiv.org/abs/hep-th/9905111}{{\ttfamily arXiv:hep-th/9905111
  [hep-th]}}.

\bibitem{Witten:1998qj}
E.~Witten, ``{Anti-de Sitter space and holography},''
  \href{http://dx.doi.org/10.4310/ATMP.1998.v2.n2.a2}{{\em Adv. Theor. Math.
  Phys.} {\bfseries 2} (1998) 253--291},
\href{http://arxiv.org/abs/hep-th/9802150}{{\ttfamily arXiv:hep-th/9802150
  [hep-th]}}.

\bibitem{Hartnoll:2016apf}
S.~A. Hartnoll, A.~Lucas, and S.~Sachdev, ``{Holographic quantum matter},''
\href{http://arxiv.org/abs/1612.07324}{{\ttfamily arXiv:1612.07324 [hep-th]}}.

\bibitem{Hartnoll:2008vx}
S.~A. Hartnoll, C.~P. Herzog, and G.~T. Horowitz, ``{Building a Holographic
  Superconductor},''
  \href{http://dx.doi.org/10.1103/PhysRevLett.101.031601}{{\em Phys. Rev.
  Lett.} {\bfseries 101} (2008) 031601},
\href{http://arxiv.org/abs/0803.3295}{{\ttfamily arXiv:0803.3295 [hep-th]}}.

\bibitem{Gubser:2008px}
S.~S. Gubser, ``{Breaking an Abelian gauge symmetry near a black hole
  horizon},'' \href{http://dx.doi.org/10.1103/PhysRevD.78.065034}{{\em Phys.
  Rev.} {\bfseries D78} (2008) 065034},
\href{http://arxiv.org/abs/0801.2977}{{\ttfamily arXiv:0801.2977 [hep-th]}}.

\bibitem{Nakamura:2009tf}
S.~Nakamura, H.~Ooguri, and C.-S. Park, ``{Gravity Dual of Spatially Modulated
  Phase},'' \href{http://dx.doi.org/10.1103/PhysRevD.81.044018}{{\em Phys.
  Rev.} {\bfseries D81} (2010) 044018},
\href{http://arxiv.org/abs/0911.0679}{{\ttfamily arXiv:0911.0679 [hep-th]}}.

\bibitem{Donos:2011bh}
A.~Donos and J.~P. Gauntlett, ``{Holographic striped phases},''
  \href{http://dx.doi.org/10.1007/JHEP08(2011)140}{{\em JHEP} {\bfseries 08}
  (2011) 140},
\href{http://arxiv.org/abs/1106.2004}{{\ttfamily arXiv:1106.2004 [hep-th]}}.

\bibitem{Donos:2011ff}
A.~Donos and J.~P. Gauntlett, ``{Holographic helical superconductors},''
  \href{http://dx.doi.org/10.1007/JHEP12(2011)091}{{\em JHEP} {\bfseries 12}
  (2011) 091},
\href{http://arxiv.org/abs/1109.3866}{{\ttfamily arXiv:1109.3866 [hep-th]}}.

\bibitem{Donos:2011qt}
A.~Donos, J.~P. Gauntlett, and C.~Pantelidou, ``{Spatially modulated
  instabilities of magnetic black branes},''
  \href{http://dx.doi.org/10.1007/JHEP01(2012)061}{{\em JHEP} {\bfseries 01}
  (2012) 061},
\href{http://arxiv.org/abs/1109.0471}{{\ttfamily arXiv:1109.0471 [hep-th]}}.

\bibitem{Son:2007vk}
D.~T. Son and A.~O. Starinets, ``{Viscosity, Black Holes, and Quantum Field
  Theory},''
  \href{http://dx.doi.org/10.1146/annurev.nucl.57.090506.123120}{{\em Ann. Rev.
  Nucl. Part. Sci.} {\bfseries 57} (2007) 95--118},
\href{http://arxiv.org/abs/0704.0240}{{\ttfamily arXiv:0704.0240 [hep-th]}}.

\bibitem{Bhattacharyya:2008jc}
S.~Bhattacharyya, V.~E. Hubeny, S.~Minwalla, and M.~Rangamani, ``{Nonlinear
  Fluid Dynamics from Gravity},''
  \href{http://dx.doi.org/10.1088/1126-6708/2008/02/045}{{\em JHEP} {\bfseries
  0802} (2008) 045},
\href{http://arxiv.org/abs/0712.2456}{{\ttfamily arXiv:0712.2456 [hep-th]}}.

\bibitem{Haack:2008cp}
M.~Haack and A.~Yarom, ``{Nonlinear viscous hydrodynamics in various dimensions
  using AdS/CFT},'' \href{http://dx.doi.org/10.1088/1126-6708/2008/10/063}{{\em
  JHEP} {\bfseries 10} (2008) 063},
\href{http://arxiv.org/abs/0806.4602}{{\ttfamily arXiv:0806.4602 [hep-th]}}.

\bibitem{Erdmenger:2008rm}
J.~Erdmenger, M.~Haack, M.~Kaminski, and A.~Yarom, ``{Fluid dynamics of
  R-charged black holes},''
  \href{http://dx.doi.org/10.1088/1126-6708/2009/01/055}{{\em JHEP} {\bfseries
  01} (2009) 055},
\href{http://arxiv.org/abs/0809.2488}{{\ttfamily arXiv:0809.2488 [hep-th]}}.

\bibitem{Banerjee:2008th}
N.~Banerjee, J.~Bhattacharya, S.~Bhattacharyya, S.~Dutta, R.~Loganayagam, and
  P.~Surowka, ``{Hydrodynamics from charged black branes},''
  \href{http://dx.doi.org/10.1007/JHEP01(2011)094}{{\em JHEP} {\bfseries 01}
  (2011) 094}, \href{http://arxiv.org/abs/0809.2596}{{\ttfamily arXiv:0809.2596
  [hep-th]}}.

\bibitem{Baggioli:2022pyb}
M.~Baggioli and B.~Gout\'eraux, ``{Effective and holographic theories of
  strongly-correlated phases of matter with broken translations},''
  \href{http://arxiv.org/abs/2203.03298}{{\ttfamily arXiv:2203.03298
  [hep-th]}}.

\bibitem{Herzog:2011ec}
C.~P. Herzog, N.~Lisker, P.~Surowka, and A.~Yarom, ``{Transport in holographic
  superfluids},'' \href{http://dx.doi.org/10.1007/JHEP08(2011)052}{{\em JHEP}
  {\bfseries 08} (2011) 052},
\href{http://arxiv.org/abs/1101.3330}{{\ttfamily arXiv:1101.3330 [hep-th]}}.

\bibitem{Bhattacharya:2011tra}
J.~Bhattacharya, S.~Bhattacharyya, S.~Minwalla, and A.~Yarom, ``{A Theory of
  first order dissipative superfluid dynamics},''
  \href{http://dx.doi.org/10.1007/JHEP05(2014)147}{{\em JHEP} {\bfseries 05}
  (2014) 147},
\href{http://arxiv.org/abs/1105.3733}{{\ttfamily arXiv:1105.3733 [hep-th]}}.

\bibitem{Bhattacharya:2011eea}
J.~Bhattacharya, S.~Bhattacharyya, and S.~Minwalla, ``{Dissipative Superfluid
  dynamics from gravity},''
  \href{http://dx.doi.org/10.1007/JHEP04(2011)125}{{\em JHEP} {\bfseries 1104}
  (2011) 125},
\href{http://arxiv.org/abs/1101.3332}{{\ttfamily arXiv:1101.3332 [hep-th]}}.

\bibitem{Donos:2022xfd}
A.~Donos and C.~Pantelidou, ``{Higgs/Amplitude Mode Dynamics From
  Holography},'' \href{http://arxiv.org/abs/2205.06294}{{\ttfamily
  arXiv:2205.06294 [hep-th]}}.

\bibitem{Amado:2009ts}
I.~Amado, M.~Kaminski, and K.~Landsteiner, ``{Hydrodynamics of Holographic
  Superconductors},''
  \href{http://dx.doi.org/10.1088/1126-6708/2009/05/021}{{\em JHEP} {\bfseries
  05} (2009) 021},
\href{http://arxiv.org/abs/0903.2209}{{\ttfamily arXiv:0903.2209 [hep-th]}}.

\bibitem{Amado:2013xya}
I.~Amado, D.~Arean, A.~Jimenez-Alba, K.~Landsteiner, L.~Melgar, and I.~S.
  Landea, ``{Holographic Type II Goldstone bosons},''
  \href{http://dx.doi.org/10.1007/JHEP07(2013)108}{{\em JHEP} {\bfseries 07}
  (2013) 108},
\href{http://arxiv.org/abs/1302.5641}{{\ttfamily arXiv:1302.5641 [hep-th]}}.

\bibitem{Bhaseen:2012gg}
M.~J. Bhaseen, J.~P. Gauntlett, B.~D. Simons, J.~Sonner, and T.~Wiseman,
  ``{Holographic Superfluids and the Dynamics of Symmetry Breaking},''
  \href{http://dx.doi.org/10.1103/PhysRevLett.110.015301}{{\em Phys. Rev.
  Lett.} {\bfseries 110} no.~1, (2013) 015301},
  \href{http://arxiv.org/abs/1207.4194}{{\ttfamily arXiv:1207.4194 [hep-th]}}.

\bibitem{Arean:2021tks}
D.~Arean, M.~Baggioli, S.~Grieninger, and K.~Landsteiner, ``{A holographic
  superfluid symphony},'' \href{http://dx.doi.org/10.1007/JHEP11(2021)206}{{\em
  JHEP} {\bfseries 11} (2021) 206},
  \href{http://arxiv.org/abs/2107.08802}{{\ttfamily arXiv:2107.08802
  [hep-th]}}.

\bibitem{Herzog:2010vz}
C.~P. Herzog, ``{An Analytic Holographic Superconductor},''
  \href{http://dx.doi.org/10.1103/PhysRevD.81.126009}{{\em Phys. Rev.}
  {\bfseries D81} (2010) 126009},
\href{http://arxiv.org/abs/1003.3278}{{\ttfamily arXiv:1003.3278 [hep-th]}}.

\bibitem{Donos:2021pkk}
A.~Donos, P.~Kailidis, and C.~Pantelidou, ``{Dissipation in holographic
  superfluids},'' \href{http://dx.doi.org/10.1007/JHEP09(2021)134}{{\em JHEP}
  {\bfseries 09} (2021) 134}, \href{http://arxiv.org/abs/2107.03680}{{\ttfamily
  arXiv:2107.03680 [hep-th]}}.

\bibitem{Crnkovic:1986ex}
C.~Crnkovic and E.~Witten, ``{Covariant description of canonical formalism in
  geometrical theories},''.

\bibitem{Iqbal:2008by}
N.~Iqbal and H.~Liu, ``{Universality of the hydrodynamic limit in AdS/CFT and
  the membrane paradigm},''
  \href{http://dx.doi.org/10.1103/PhysRevD.79.025023}{{\em Phys.Rev.}
  {\bfseries D79} (2009) 025023},
\href{http://arxiv.org/abs/0809.3808}{{\ttfamily arXiv:0809.3808 [hep-th]}}.

\bibitem{Eling:2011ms}
C.~Eling and Y.~Oz, ``{A Novel Formula for Bulk Viscosity from the Null Horizon
  Focusing Equation},'' \href{http://dx.doi.org/10.1007/JHEP06(2011)007}{{\em
  JHEP} {\bfseries 06} (2011) 007},
  \href{http://arxiv.org/abs/1103.1657}{{\ttfamily arXiv:1103.1657 [hep-th]}}.

\bibitem{Donos:2022uea}
A.~Donos, P.~Kailidis, and C.~Pantelidou, ``{Holographic Dissipation from the
  Symplectic Current},'' \href{http://arxiv.org/abs/2208.05911}{{\ttfamily
  arXiv:2208.05911 [hep-th]}}.

\bibitem{Skenderis:2002wp}
K.~Skenderis, ``{Lecture notes on holographic renormalization},''
  \href{http://dx.doi.org/10.1088/0264-9381/19/22/306}{{\em Class.Quant.Grav.}
  {\bfseries 19} (2002) 5849--5876},
\href{http://arxiv.org/abs/hep-th/0209067}{{\ttfamily arXiv:hep-th/0209067
  [hep-th]}}.

\bibitem{Papadimitriou:2005ii}
I.~Papadimitriou and K.~Skenderis, ``{Thermodynamics of asymptotically locally
  AdS spacetimes},''
  \href{http://dx.doi.org/10.1088/1126-6708/2005/08/004}{{\em JHEP} {\bfseries
  0508} (2005) 004},
\href{http://arxiv.org/abs/hep-th/0505190}{{\ttfamily arXiv:hep-th/0505190
  [hep-th]}}.

\bibitem{Donos:2014cya}
A.~Donos and J.~P. Gauntlett, ``{Thermoelectric DC conductivities from black
  hole horizons},'' \href{http://dx.doi.org/10.1007/JHEP11(2014)081}{{\em JHEP}
  {\bfseries 1411} (2014) 081},
\href{http://arxiv.org/abs/1406.4742}{{\ttfamily arXiv:1406.4742 [hep-th]}}.

\bibitem{Donos:2015gia}
A.~Donos and J.~P. Gauntlett, ``{Navier-Stokes Equations on Black Hole Horizons
  and DC Thermoelectric Conductivity},''
  \href{http://dx.doi.org/10.1103/PhysRevD.92.121901}{{\em Phys. Rev.}
  {\bfseries D92} no.~12, (2015) 121901},
\href{http://arxiv.org/abs/1506.01360}{{\ttfamily arXiv:1506.01360 [hep-th]}}.

\bibitem{Horowitz:2009ij}
G.~T. Horowitz and M.~M. Roberts, ``{Zero Temperature Limit of Holographic
  Superconductors},''
  \href{http://dx.doi.org/10.1088/1126-6708/2009/11/015}{{\em JHEP} {\bfseries
  0911} (2009) 015},
\href{http://arxiv.org/abs/0908.3677}{{\ttfamily arXiv:0908.3677 [hep-th]}}.

\bibitem{Gubser:2009gp}
S.~S. Gubser, S.~S. Pufu, and F.~D. Rocha, ``{Quantum critical superconductors
  in string theory and M-theory},''
  \href{http://dx.doi.org/10.1016/j.physletb.2009.12.017}{{\em Phys. Lett.}
  {\bfseries B683} (2010) 201--204},
\href{http://arxiv.org/abs/0908.0011}{{\ttfamily arXiv:0908.0011 [hep-th]}}.

\bibitem{Gouteraux:2012yr}
B.~Gouteraux and E.~Kiritsis, ``{Quantum critical lines in holographic phases
  with (un)broken symmetry},''
  \href{http://dx.doi.org/10.1007/JHEP04(2013)053}{{\em JHEP} {\bfseries 04}
  (2013) 053}, \href{http://arxiv.org/abs/1212.2625}{{\ttfamily arXiv:1212.2625
  [hep-th]}}.

\end{thebibliography}\endgroup
\end{document}